\documentclass[journal,dvipsnames]{IEEEtran}
\usepackage{graphicx,url}
\graphicspath{ {./images/} }
\usepackage{amssymb,amsmath,mathtools}
\usepackage{tikz,pgfplots}
\usepackage{pgfplots}
\usepackage{pgfplotstable}
\usetikzlibrary{pgfplots.groupplots}
\usepgfplotslibrary{colorbrewer}
\pgfplotsset{compat = 1.15, cycle list/Set1-8}
\usetikzlibrary{pgfplots.statistics, pgfplots.colorbrewer}
\usetikzlibrary{pgfplots.groupplots}
\usetikzlibrary{shapes.geometric,backgrounds,patterns, trees}
\usetikzlibrary{3d,decorations.text,shapes.arrows,positioning,fit,backgrounds}
\usetikzlibrary{positioning, decorations.pathmorphing, shapes}
\usetikzlibrary{decorations.pathreplacing}
\usetikzlibrary{shapes.geometric,backgrounds,patterns, trees}
\usetikzlibrary{spy}
\usetikzlibrary{arrows.meta,
                bending,
                intersections,
                quotes,
                shapes.geometric}
              \usetikzlibrary{automata, positioning}
              \usepgfplotslibrary{fillbetween}
\usetikzlibrary{shapes,arrows}
\usetikzlibrary{arrows.meta}
\usetikzlibrary{positioning}
\tikzset{set/.style={draw,circle,inner sep=0pt,align=center}}
\usetikzlibrary{automata, positioning}
  \usetikzlibrary{shapes,shadows}
  \tikzstyle{abstractbox} = [draw=black, fill=white, rectangle,
  inner sep=10pt, style=rounded corners, drop shadow={fill=black,
  opacity=1}]
\tikzstyle{abstracttitle} =[fill=white]
\usetikzlibrary{calc,positioning,shapes.geometric}
\usetikzlibrary{arrows.meta,arrows}
\usetikzlibrary{matrix}

\colorlet{myRed}{red!20}
\tikzset{
  rows/.style 2 args={/utils/temp/.style={row ##1/.append style={nodes={#2}}},
    /utils/temp/.list={#1}},
  columns/.style 2 args={/utils/temp/.style={column ##1/.append style={nodes={#2}}},
    /utils/temp/.list={#1}}}
\usetikzlibrary{backgrounds,calc,shadings,shapes.arrows,shapes.symbols,shadows}
\definecolor{switch}{HTML}{006996}
\pgfkeys{/pgf/.cd,
  parallelepiped offset x/.initial=2mm,
  parallelepiped offset y/.initial=2mm
}
\pgfdeclareshape{parallelepiped}
{
  \inheritsavedanchors[from=rectangle] 
  \inheritanchorborder[from=rectangle]
  \inheritanchor[from=rectangle]{north}
  \inheritanchor[from=rectangle]{north west}
  \inheritanchor[from=rectangle]{north east}
  \inheritanchor[from=rectangle]{center}
  \inheritanchor[from=rectangle]{west}
  \inheritanchor[from=rectangle]{east}
  \inheritanchor[from=rectangle]{mid}
  \inheritanchor[from=rectangle]{mid west}
  \inheritanchor[from=rectangle]{mid east}
  \inheritanchor[from=rectangle]{base}
  \inheritanchor[from=rectangle]{base west}
  \inheritanchor[from=rectangle]{base east}
  \inheritanchor[from=rectangle]{south}
  \inheritanchor[from=rectangle]{south west}
  \inheritanchor[from=rectangle]{south east}
  \backgroundpath{
    \southwest \pgf@xa=\pgf@x \pgf@ya=\pgf@y
    \northeast \pgf@xb=\pgf@x \pgf@yb=\pgf@y
    \pgfmathsetlength\pgfutil@tempdima{\pgfkeysvalueof{/pgf/parallelepiped
      offset x}}
    \pgfmathsetlength\pgfutil@tempdimb{\pgfkeysvalueof{/pgf/parallelepiped
      offset y}}
    \def\ppd@offset{\pgfpoint{\pgfutil@tempdima}{\pgfutil@tempdimb}}
    \pgfpathmoveto{\pgfqpoint{\pgf@xa}{\pgf@ya}}
    \pgfpathlineto{\pgfqpoint{\pgf@xb}{\pgf@ya}}
    \pgfpathlineto{\pgfqpoint{\pgf@xb}{\pgf@yb}}
    \pgfpathlineto{\pgfqpoint{\pgf@xa}{\pgf@yb}}
    \pgfpathclose
    \pgfpathmoveto{\pgfqpoint{\pgf@xb}{\pgf@ya}}
    \pgfpathlineto{\pgfpointadd{\pgfpoint{\pgf@xb}{\pgf@ya}}{\ppd@offset}}
    \pgfpathlineto{\pgfpointadd{\pgfpoint{\pgf@xb}{\pgf@yb}}{\ppd@offset}}
    \pgfpathlineto{\pgfpointadd{\pgfpoint{\pgf@xa}{\pgf@yb}}{\ppd@offset}}
    \pgfpathlineto{\pgfqpoint{\pgf@xa}{\pgf@yb}}
    \pgfpathmoveto{\pgfqpoint{\pgf@xb}{\pgf@yb}}
    \pgfpathlineto{\pgfpointadd{\pgfpoint{\pgf@xb}{\pgf@yb}}{\ppd@offset}}
  }
}

\makeatletter
\tikzset{anchor/.append code=\let\tikz@auto@anchor\relax,
  add font/.code=%
    \expandafter\def\expandafter\tikz@textfont\expandafter{\tikz@textfont#1},
  left delimiter/.style 2 args={append after command={\tikz@delimiter{south east}
    {south west}{every delimiter,every left delimiter,#2}{south}{north}{#1}{.}{\pgf@y}}}}
\tikzstyle{sms} = [rectangle callout, draw,very thick, rounded corners, minimum height=20pt]
\makeatletter
\tikzset{anchor/.append code=\let\tikz@auto@anchor\relax,
  add font/.code=%
    \expandafter\def\expandafter\tikz@textfont\expandafter{\tikz@textfont#1},
  left delimiter/.style 2 args={append after command={\tikz@delimiter{south east}
    {south west}{every delimiter,every left delimiter,#2}{south}{north}{#1}{.}{\pgf@y}}}}
\tikzstyle{sms} = [rectangle callout, draw,very thick, rounded corners, minimum height=20pt]
\usetikzlibrary{positioning,calc}
\tikzstyle{block} = [rectangle, draw,
text width=10.5em, text centered, rounded corners, minimum height=4em]
\tikzstyle{line} = [draw, -latex]
\tikzset{l3 switch/.style={
    parallelepiped,fill=switch, draw=white,
    minimum width=0.75cm,
    minimum height=0.75cm,
    parallelepiped offset x=1.75mm,
    parallelepiped offset y=1.25mm,
    path picture={
      \node[fill=white,
        circle,
        minimum size=6pt,
        inner sep=0pt,
        append after command={
          \pgfextra{
            \foreach \angle in {0,45,...,360}
            \draw[-latex,fill=white] (\tikzlastnode.\angle)--++(\angle:2.25mm);
          }
        }
      ]
       at ([xshift=-0.75mm,yshift=-0.5mm]path picture bounding box.center){};
    }
  },
  ports/.style={
    line width=0.3pt,
    top color=gray!20,
    bottom color=gray!80
  },
  rack switch/.style={
    parallelepiped,fill=white, draw,
    minimum width=1.25cm,
    minimum height=0.25cm,
    parallelepiped offset x=2mm,
    parallelepiped offset y=1.25mm,
    xscale=-1,
    path picture={
      \draw[top color=gray!5,bottom color=gray!40]
      (path picture bounding box.south west) rectangle
      (path picture bounding box.north east);
      \coordinate (A-west) at ([xshift=-0.2cm]path picture bounding box.west);
      \coordinate (A-center) at ($(path picture bounding box.center)!0!(path
        picture bounding box.south)$);
      \foreach \x in {0.275,0.525,0.775}{
        \draw[ports]([yshift=-0.05cm]$(A-west)!\x!(A-center)$)
          rectangle +(0.1,0.05);
        \draw[ports]([yshift=-0.125cm]$(A-west)!\x!(A-center)$)
          rectangle +(0.1,0.05);
       }
      \coordinate (A-east) at (path picture bounding box.east);
      \foreach \x in {0.085,0.21,0.335,0.455,0.635,0.755,0.875,1}{
        \draw[ports]([yshift=-0.1125cm]$(A-east)!\x!(A-center)$)
          rectangle +(0.05,0.1);
      }
    }
  },
  server/.style={
    parallelepiped,
    fill=white, draw,
    minimum width=0.35cm,
    minimum height=0.75cm,
    parallelepiped offset x=3mm,
    parallelepiped offset y=2mm,
    xscale=-1,
    path picture={
      \draw[top color=gray!5,bottom color=gray!40]
      (path picture bounding box.south west) rectangle
      (path picture bounding box.north east);
      \coordinate (A-center) at ($(path picture bounding box.center)!0!(path
        picture bounding box.south)$);
      \coordinate (A-west) at ([xshift=-0.575cm]path picture bounding box.west);
      \draw[ports]([yshift=0.1cm]$(A-west)!0!(A-center)$)
        rectangle +(0.2,0.065);
      \draw[ports]([yshift=0.01cm]$(A-west)!0.085!(A-center)$)
        rectangle +(0.15,0.05);
      \fill[black]([yshift=-0.35cm]$(A-west)!-0.1!(A-center)$)
        rectangle +(0.235,0.0175);
      \fill[black]([yshift=-0.385cm]$(A-west)!-0.1!(A-center)$)
        rectangle +(0.235,0.0175);
      \fill[black]([yshift=-0.42cm]$(A-west)!-0.1!(A-center)$)
        rectangle +(0.235,0.0175);
    }
  },
}

\usetikzlibrary{calc, shadings, shadows, shapes.arrows}
\tikzset{cross/.style={cross out, draw=black, minimum size=2*(#1-\pgflinewidth), inner sep=0pt, outer sep=0pt},
cross/.default={1pt}}
\tikzset{%
  interface/.style={draw, rectangle, rounded corners, font=\LARGE\sffamily},
  ethernet/.style={interface, fill=yellow!50},
  serial/.style={interface, fill=green!70},
  speed/.style={sloped, anchor=south, font=\large\sffamily},
  route/.style={draw, shape=single arrow, single arrow head extend=4mm,
    minimum height=1.7cm, minimum width=3mm, white, fill=switch!20,
    drop shadow={opacity=.8, fill=switch}, font=\tiny}
}

\usepackage{float}
\definecolor{bluetwo}{RGB}{189, 213, 234}
\definecolor{bluethree}{RGB}{165, 193, 224}
\definecolor{bluefour}{RGB}{141, 169, 200}

\usepackage{amsthm}
\theoremstyle{plain}
\newtheorem{proposition}{Proposition}
\newtheorem{definition}{Definition}

\newcommand\norm[1]{\lVert#1\rVert}
\newcommand\numeq[1]%
{\stackrel{\scriptscriptstyle(\mkern-1.5mu#1\mkern-1.5mu)}{=}}
\newcommand\numeqq[1]%
{\stackrel{\scriptscriptstyle(\mkern-1.5mu#1\mkern-1.5mu)}{\triangleq}}
\newcommand\numleq[1]%
{\stackrel{\scriptscriptstyle(\mkern-1.5mu#1\mkern-1.5mu)}{\leq}}
\newcommand\numgeq[1]%
{\stackrel{\scriptscriptstyle(\mkern-1.5mu#1\mkern-1.5mu)}{\geq}}
\newcommand\numimp[1]%
{\stackrel{\scriptscriptstyle(\mkern-1.5mu#1\mkern-1.5mu)}{\implies}}
\newtheorem{remark}{Remark}

\usepackage{booktabs}
\usepackage{colortbl}
\definecolor{lightgray}{gray}{0.9}
\allowdisplaybreaks

\definecolor{lightgray}{gray}{0.9}
\definecolor{lightgray}{gray}{0.9}
\definecolor{lightgreen}{rgb}{0.88, 1, 0.88}
\definecolor{lightred}{rgb}{1, 0.88, 0.88}
\definecolor{lightblue}{rgb}{0.88, 0.94, 1}
\definecolor{lightorange}{rgb}{1, 0.94, 0.88}

\usepackage{bbm}
\usepackage{bm}

\DeclareMathOperator*{\argmax}{arg\,max}

\DeclareMathOperator*{\maximize}{maximize}

\definecolor{gray2}{HTML}{ededed}
\definecolor{gray3}{HTML}{F5F5F5}
\definecolor{RoyalAzure}{rgb}{0.0, 0.22, 0.66}
\definecolor{lightgray}{gray}{0.9}
\definecolor{lightgray}{gray}{0.9}
\definecolor{lightgreen}{rgb}{0.88, 1, 0.88}
\definecolor{lightred}{rgb}{1, 0.88, 0.88}
\definecolor{lightblue}{rgb}{0.88, 0.94, 1}
\definecolor{lightorange}{rgb}{1, 0.94, 0.88}

\makeatletter
\tikzset{
    database/.style={
        path picture={
            \draw (0, 1.5*\database@segmentheight) circle [x radius=\database@radius,y radius=\database@aspectratio*\database@radius];
            \draw (-\database@radius, 0.5*\database@segmentheight) arc [start angle=180,end angle=360,x radius=\database@radius, y radius=\database@aspectratio*\database@radius];
            \draw (-\database@radius,-0.5*\database@segmentheight) arc [start angle=180,end angle=360,x radius=\database@radius, y radius=\database@aspectratio*\database@radius];
            \draw (-\database@radius,1.5*\database@segmentheight) -- ++(0,-3*\database@segmentheight) arc [start angle=180,end angle=360,x radius=\database@radius, y radius=\database@aspectratio*\database@radius] -- ++(0,3*\database@segmentheight);
        },
        minimum width=2*\database@radius + \pgflinewidth,
        minimum height=3*\database@segmentheight + 2*\database@aspectratio*\database@radius + \pgflinewidth,
    },
    database segment height/.store in=\database@segmentheight,
    database radius/.store in=\database@radius,
    database aspect ratio/.store in=\database@aspectratio,
    database segment height=0.1cm,
    database radius=0.25cm,
    database aspect ratio=0.35,
  }
\makeatother

\usetikzlibrary{backgrounds}
\usetikzlibrary{patterns}
\pgfmathdeclarefunction{gauss}{3}{%
  \pgfmathparse{1/(#3*sqrt(2*pi))*exp(-((#1-#2)^2)/(2*#3^2))}%
}

\usetikzlibrary{spy}
\usetikzlibrary{arrows.meta,
                bending,
                intersections,
                quotes,
                shapes.geometric}
              \usetikzlibrary{automata, positioning}
              \usepgfplotslibrary{fillbetween}

\tikzset{%
  interface/.style={draw, rectangle, rounded corners, font=\LARGE\sffamily},
  ethernet/.style={interface, fill=yellow!50},
  serial/.style={interface, fill=green!70},
  speed/.style={sloped, anchor=south, font=\large\sffamily},
  route/.style={draw, shape=single arrow, single arrow head extend=4mm,
    minimum height=1.7cm, minimum width=3mm, white, fill=switch!20,
    drop shadow={opacity=.8, fill=switch}, font=\tiny}
}
\definecolor{switch}{HTML}{006996}

\tikzset{l3 switch/.style={
    parallelepiped,fill=switch, draw=white,
    minimum width=0.75cm,
    minimum height=0.75cm,
    parallelepiped offset x=1.75mm,
    parallelepiped offset y=1.25mm,
    path picture={
      \node[fill=white,
      circle,
      minimum size=6pt,
      inner sep=0pt,
      append after command={
        \pgfextra{
          \foreach \angle in {0,45,...,360}
          \draw[-latex,fill=white] (\tikzlastnode.\angle)--++(\angle:2.25mm);
        }
      }
      ]
      at ([xshift=-0.75mm,yshift=-0.5mm]path picture bounding box.center){};
    }
  },
  ports/.style={
    line width=0.3pt,
    top color=gray!20,
    bottom color=gray!80
  },
  rack switch/.style={
    parallelepiped,fill=white, draw,
    minimum width=1.25cm,
    minimum height=0.25cm,
    parallelepiped offset x=2mm,
    parallelepiped offset y=1.25mm,
    xscale=-1,
    path picture={
      \draw[top color=gray!5,bottom color=gray!40]
      (path picture bounding box.south west) rectangle
      (path picture bounding box.north east);
      \coordinate (A-west) at ([xshift=-0.2cm]path picture bounding box.west);
      \coordinate (A-center) at ($(path picture bounding box.center)!0!(path
      picture bounding box.south)$);
      \foreach \x in {0.275,0.525,0.775}{
        \draw[ports]([yshift=-0.05cm]$(A-west)!\x!(A-center)$)
        rectangle +(0.1,0.05);
        \draw[ports]([yshift=-0.125cm]$(A-west)!\x!(A-center)$)
        rectangle +(0.1,0.05);
      }
      \coordinate (A-east) at (path picture bounding box.east);
      \foreach \x in {0.085,0.21,0.335,0.455,0.635,0.755,0.875,1}{
        \draw[ports]([yshift=-0.1125cm]$(A-east)!\x!(A-center)$)
        rectangle +(0.05,0.1);
      }
    }
  },
  server/.style={
    parallelepiped,
    fill=white, draw,
    minimum width=0.35cm,
    minimum height=0.75cm,
    parallelepiped offset x=3mm,
    parallelepiped offset y=2mm,
    xscale=-1,
    path picture={
      \draw[top color=gray!5,bottom color=gray!40]
      (path picture bounding box.south west) rectangle
      (path picture bounding box.north east);
      \coordinate (A-center) at ($(path picture bounding box.center)!0!(path
      picture bounding box.south)$);
      \coordinate (A-west) at ([xshift=-0.575cm]path picture bounding box.west);
      \draw[ports]([yshift=0.1cm]$(A-west)!0!(A-center)$)
      rectangle +(0.2,0.065);
      \draw[ports]([yshift=0.01cm]$(A-west)!0.085!(A-center)$)
      rectangle +(0.15,0.05);
      \fill[black]([yshift=-0.35cm]$(A-west)!-0.1!(A-center)$)
      rectangle +(0.235,0.0175);
      \fill[black]([yshift=-0.385cm]$(A-west)!-0.1!(A-center)$)
      rectangle +(0.235,0.0175);
      \fill[black]([yshift=-0.42cm]$(A-west)!-0.1!(A-center)$)
      rectangle +(0.235,0.0175);
    }
  },
}

\makeatletter
\pgfdeclareradialshading[tikz@ball]{cloud}{\pgfpoint{-0.275cm}{0.4cm}}{%
  color(0cm)=(tikz@ball!75!white);
  color(0.1cm)=(tikz@ball!85!white);
  color(0.2cm)=(tikz@ball!95!white);
  color(0.7cm)=(tikz@ball!89!black);
  color(1cm)=(tikz@ball!75!black)
}
\tikzoption{cloud color}{\pgfutil@colorlet{tikz@ball}{#1}%
  \def\tikz@shading{cloud}\tikz@addmode{\tikz@mode@shadetrue}}
\makeatother

\tikzset{my cloud/.style={
     cloud, draw, aspect=2,
     cloud color={gray!5!white}
  }
}

\tikzset{
  mybackground18/.style={execute at end picture={
      \begin{scope}[on background layer]
        \draw[black, fill=gray3, rounded corners=2.5ex] (current bounding box.south west)
        rectangle (current bounding box.north east);
        \node[draw,fill=white,ellipse,anchor=west,inner sep=1pt,minimum width=4ex] at (current bounding box.north
        west){#1};
      \end{scope}
    }}
}
\tikzset{
  mybackground48/.style={execute at end picture={
      \begin{scope}[on background layer]
        \draw[black, fill=gray3, rounded corners=2.5ex] (current bounding box.south west)
        rectangle (current bounding box.north east);
      \end{scope}
    }}
}

\usepackage{listofitems} 
\usetikzlibrary{arrows.meta} 
\usepackage[outline]{contour} 
\contourlength{1.4pt}

\colorlet{myred}{red!80!black}
\colorlet{myblue}{blue!80!black}
\colorlet{mygreen}{green!60!black}
\colorlet{myorange}{orange!70!red!60!black}
\colorlet{mydarkred}{red!30!black}
\colorlet{mydarkblue}{blue!40!black}
\colorlet{mydarkgreen}{green!30!black}

\tikzset{
  >=latex, 
  node/.style={thick,circle,draw=myblue,minimum size=22,inner sep=0.5,outer sep=0.6},
  node in/.style={node,black!20!black,draw=mygreen!30!black,fill=black!20},
  node hidden/.style={node,black!20!black,draw=myblue!30!black,fill=black!20},
  node convol/.style={node,black!20!black,draw=myorange!30!black,fill=black!20},
  node out/.style={node,red!20!black,draw=myred!30!black,fill=black!20},
  connect/.style={thick,Blue!100}, 
  connect arrow/.style={-{Latex[length=4,width=3.5]},thick,mydarkblue,shorten <=0.5,shorten >=1},
  node 1/.style={node in}, 
  node 2/.style={node hidden},
  node 3/.style={node out}
}

\usepackage{makecell}
\usepackage[most]{tcolorbox}

\usepackage[titlenumbered,ruled,linesnumbered]{algorithm2e}
\SetAlCapNameFnt{\footnotesize}
\SetAlCapFnt{\footnotesize}

\usepackage{pifont}
\newcommand{\cmark}{\textcolor{OliveGreen}{\ding{51}}} 
\newcommand{\xmark}{\textcolor{Red}{\ding{55}}}

\newcommand{\legendSymbol}[1]{%
    \tikz[baseline=-0.5ex]{
      \draw[RoyalAzure,thick, solid] (0,0) -- (0.5cm,0);
      \path[RoyalAzure,thick] plot coordinates {(0.25cm,0)};      
    }%
  }

\newcommand{\legendSymboltwo}[1]{%
    \tikz[baseline=-0.5ex]{
      \draw[Red,thick, solid,densely dashdotted] (0,0) -- (0.5cm,0);
      \path[Red,thick,densely dashdotted] plot coordinates {(0.25cm,0)};      
    }%
  }
\newcommand{\legendSymbolthree}[1]{%
    \tikz[baseline=-0.5ex]{
      \draw[OliveGreen,thick, solid,dashed] (0,0) -- (0.5cm,0);
      \path[OliveGreen,thick,dashed] plot coordinates {(0.25cm,0)};      
    }%
  }  
\usepackage{mathrsfs}
\newcommand{\indep}{\perp \!\!\! \perp}
\let\oldnl\nl
\newcommand{\nonl}{\renewcommand{\nl}{\let\nl\oldnl}}
\usepackage{lettrine}
\usepackage{cite}
\definecolor{grad1}{RGB}{0,150,150}
\definecolor{grad2}{RGB}{0,100,200}
\definecolor{grad3}{RGB}{80,0,150}
\definecolor{grad4}{RGB}{180, 0, 120}
\definecolor{grad5}{RGB}{255, 80, 80}
\definecolor{grad0}{RGB}{0, 150, 80}
\definecolor{grad6}{RGB}{255, 80, 80}
\definecolor{gray11}{RGB}{45, 55, 60}

\begin{document}

\title{\textbf{C}ausal \textbf{O}nline \textbf{L}earning of Safe Regions \\in Cloud Radio Access Networks}

\author{\IEEEauthorblockN{Kim Hammar\IEEEauthorrefmark{2}, Tansu Alpcan\IEEEauthorrefmark{2}, Emil C. Lupu\IEEEauthorrefmark{3}}\\
 \IEEEauthorblockA{\IEEEauthorrefmark{2}
   Department of Electrical and Electronic Engineering, University of Melbourne, Australia\\
 } 
 \IEEEauthorblockA{\IEEEauthorrefmark{3}
   Department of Computing, Imperial College London, United Kingdom\\
 }
 Email: \{kim.hammar,tansu.alpcan\}@unimelb.edu.au, e.c.lupu@imperial.ac.uk
}

\maketitle
\begin{abstract}
Cloud radio access networks (RANs) enable cost-effective management of mobile networks by dynamically scaling their capacity on demand. However, deploying adaptive controllers to implement such dynamic scaling in operational networks is challenging due to the risk of breaching service agreements and operational constraints. To mitigate this challenge, we present a novel method for learning the safe operating region of the RAN, i.e., the set of resource allocations and network configurations for which its specification is fulfilled. The method, which we call (C)ausal (O)nline (L)earning, operates in two online phases: an inference phase and an intervention phase. In the first phase, we passively observe the RAN to infer an initial safe region via causal inference and Gaussian process regression. In the second phase, we gradually expand this region through interventional Bayesian learning. We prove that COL ensures that the learned region is safe with a specified probability and that it converges to the full safe region under standard conditions. We experimentally validate COL on a 5G testbed. The results show that COL quickly learns the safe region while incurring low operational cost and being up to $10\times$ more sample-efficient than current state-of-the-art methods for safe learning.
\end{abstract}
\IEEEpeerreviewmaketitle
\section{Introduction}
\lettrine[lines=2]{\textbf{C}}{loud} radio access networks (RANs) introduce a new paradigm in mobile network design, where network functions are virtualized and disaggregated across cloud infrastructures. Such virtualization decouples the baseband processing in the RAN from the hardware at the cell sites, which allows operators to flexibly allocate compute and radio resources across the network to support varying traffic demands \cite{9253665}. While this added flexibility can enable cost-effective network management, it also introduces new operational challenges in adaptive control and allocation of network resources \cite{6897914}.

The key challenge in operating the RAN is understanding how the RAN's \textit{control inputs} (e.g., network configurations and resource allocations) jointly influence its performance and behavior. These control inputs are often interdependent in complex ways, which means that small adjustments to a single input can trigger nonlinear or cascading effects on many other system variables, such as network latency, throughput, and availability. Consequently, configuring the RAN to meet strict service-level requirements is a major technical challenge, particularly for RANs running on shared cloud infrastructures with fluctuating workloads and varying service requirements.

\begin{figure}
  \centering
  \scalebox{0.95}{
   \includegraphics{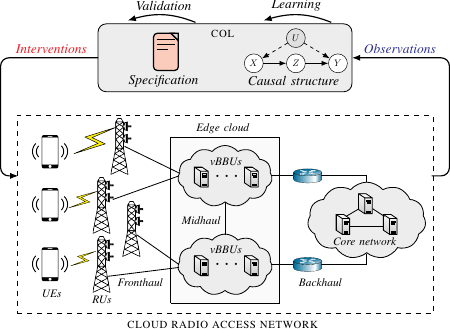}    
  }
  \caption{Causal Online Learning (COL): a method for identifying the \textit{safe region} of a cloud radio access network (RAN) where virtual baseband units (vBBUs) are distributed across an (edge) cloud infrastructure. The safe region contains the set of \textit{control inputs} (e.g., resource allocations) for which the RAN's \textit{specification} (i.e., service requirements and constraints) is fulfilled. COL exploits the causal structure of the RAN to learn the safe region from system observations, which are collected through a sequence of \textit{interventions}.}
  \label{fig:config_space}
\end{figure}

To mitigate these challenges and ensure reliable operation, it is necessary to identify the RAN's \textit{safe operating region}, i.e., the subset of control inputs for which the RAN satisfies its \textit{specification}; see Fig.~\ref{fig:config_space}. For example, the specification may require that processing delays remain below certain thresholds to support ultra-reliable low-latency communication (URLLC) \cite{10679265}. We say that a control input is \textit{safe} if it does not cause the RAN to violate its specification. Knowing the region of such control inputs provides two key benefits: (\textit{i}) it supports operators in configuring the RAN reliably; and (\textit{ii}) it enables the safe optimization and automation of control inputs to meet performance objectives. As a consequence, it provides a foundation for autonomous management paradigms, such as zero-touch management \cite{coronado2022zero} and intent-based networking \cite{clemm2022rfc}. 

Current approaches for identifying safe regions include methods from safe control (see e.g., \cite{10.5555/1717907}), safe reinforcement learning (see e.g., \cite{safe_rl_survey}), and safe Bayesian optimization (see e.g., \cite{pmlr-v37-sui15}). However, these methods are unsuitable for complex RANs for the following reasons. First, traditional control methods assume knowledge of the target system's dynamics (i.e., equations that govern the system's evolution over time), which is unrealistic for RANs. Second, most safe reinforcement learning methods are designed for offline learning of a safe control policy. While this approach can ensure the safety of the learned policy, it does not ensure safety during the learning process, which is required in practice. Third, safe Bayesian optimization approaches treat the target system as a black box, which ignores the rich structural properties of RANs. For example, RANs that follow the O-RAN standard are built from modular components that interact through well-defined interfaces and protocols \cite{oran-architecture-v15}. These structured interactions constrain how disturbances propagate and how control inputs take effect throughout the network. Ignoring this structure increases both the amount of data required to accurately learn the safe region and the risk of learning spurious correlations that do not reflect the true causal dynamics of the RAN \cite{pearl2000causality}.

In this paper, we address these limitations by presenting \textbf{C}ausal \textbf{O}nline \textbf{L}earning (COL), a novel method that leverages structural properties of the RAN to efficiently learn the safe region directly from \textit{system observations} (i.e., performance metrics and system logs). COL operates in two online phases: an \textit{inference phase} and an \textit{intervention phase}. In the first phase, we passively observe the RAN to infer an initial safe region via causal inference and Gaussian process (GP) regression. This region provides a safe starting point for the second phase of COL, where we gradually expand the estimated region through a sequence of controlled interventions.

During such an intervention, we set one or more control inputs to new values and evaluate whether the specification is satisfied, which allows us to explore parts of the safe region that cannot be identified through passive monitoring of the RAN. For instance, an intervention in this context may be to temporarily adjust the CPU allocation and observe how this change affects processing delays \cite{10.1145/3447993.3483266}. We select these interventions based on the uncertainty estimates provided by the GP, which allows us to focus on interventions that are most effective for reducing uncertainty about the safe region. 

We prove that COL ensures that the learned region is safe with a specified probability and that it converges to the full safe region as more system observations are collected (under standard regularity assumptions). To evaluate COL experimentally, we use it to learn the safe operating region of a cloud RAN deployed on a 5G testbed. The results show that COL quickly approximates the safe region while having low interference with system operations. Moreover, we find that COL is up to $10\times$ more sample-efficient than methods that do not exploit the causal structure of the RAN.

Our contributions can be summarized as follows:
\begin{itemize}
  \item We develop \textbf{C}ausal \textbf{O}nline \textbf{L}earning (COL), a novel method for learning the safe region of a cloud RAN. COL exploits the RAN's causal structure and estimates the safe region with Gaussian processes that are learned in two phases: an inference phase and an intervention phase.
\item We prove that the safe region learned by COL satisfies the RAN's specification with a given probability and that it converges to the full safe region as more observations are collected (under standard assumptions).
\item We validate COL experimentally on a cloud RAN in a 5G testbed. The results show that COL quickly learns the safe region and is up to $10\times$ more sample-efficient than methods that do not exploit the causal structure.
\item We release a novel dataset comprising $168$ hours of monitoring data collected from our 5G testbed, along with the code for reproducing our experiments \cite{csle_docs}. 
\end{itemize}    

\begin{figure*}
  \centering
  \scalebox{1.15}{
   \includegraphics{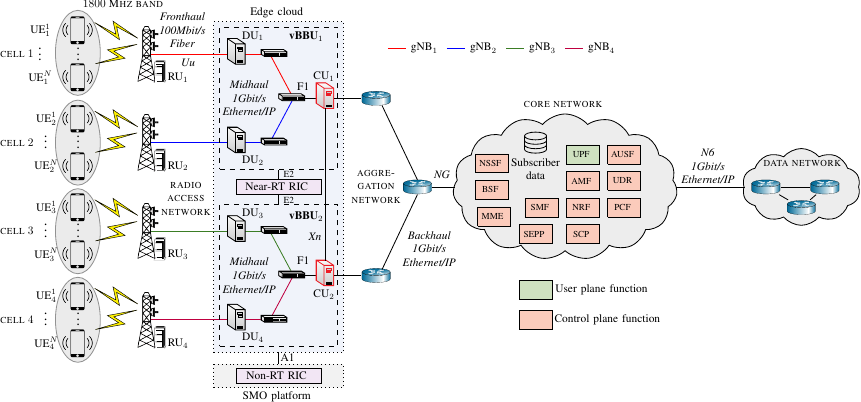}    
  }
\caption{Architecture of the target system that we study in this paper: a 5G cloud radio access network (RAN). The network follows the O-RAN architecture \cite{oran-architecture-v15} with a split distributed design in the 1800 MHz band (band 3), where four gNBs (5G base stations) provide radio access to user equipment (UE). Each gNB includes a radio unit (RU) that is connected to a distributed unit (DU) via a 100 Mbit/s fronthaul (the Uu interface). DUs are aggregated into two virtual baseband units (vBBUs), each comprising one centralized unit (CU) and two DUs. They are interconnected through a 1 Gbit/s midhaul (the F1 interface) and hosted in an edge cloud together with a near-real-time RAN intelligent controller (near-RT RIC) that interfaces with a non-RT RIC (the A1 interface), which is deployed in a service management and orchestration (SMO) platform. The vBBUs are interconnected through the Xn interface and communicate with the core network over a 1 Gbit/s backhaul (the NG interface). The core network follows a service-oriented architecture and provides (virtual) network functions, such as the access and mobility management function (AMF). It is connected to the data network (e.g., the Internet) via the N6 interface.}
  \label{fig:target_system}
\end{figure*}
\section{Related Work}
Analyzing the safety of dynamic systems is a topic of ongoing interest that has been extensively studied by control theorists and computer scientists for many decades. Traditional approaches include Lyapunov analysis \cite{10.5555/944919.944955}, reachability computations \cite{LYGEROS1999349}, robust control \cite{zhou1998essentials}, control barrier functions \cite{WIELAND2007462}, model predictive control \cite{10.5555/3164811}, and formal methods for model checking \cite{Baier2008}. Most of these techniques were originally developed for control systems with well-specified dynamics. More recently, analogous methods have been developed for machine learning systems, most notably methods based on conformal prediction theory \cite{10.5555/1062391} and certified robustness \cite{cohen2019certified}.

While these methods focus on validating the safety of systems with known dynamics, many applications involve systems where the dynamics are unknown and must be learned. This need has motivated a growing body of work on learning safe regions directly from interaction with the system, in particular approaches based on safe reinforcement learning \cite{safe_rl_survey} and safe Bayesian optimization \cite{pmlr-v37-sui15}. From a methodological perspective, these approaches can be divided into two groups: (\textit{i}) \textit{offline} methods that focus on the problem of learning a safe control policy, while not necessarily ensuring that the learning process is safe; and (\textit{ii}) \textit{online} methods that focus on ensuring safety while learning the control policy; see Table~\ref{tab:related_work}.
\begin{table}[H]
  \centering
  \scalebox{0.75}{
    \begin{tabular}{cccc} \toprule
\rowcolor{lightgray}
      {\textit{Method}} & {\textit{Online}} & {\textit{System knowledge}} & {\textit{Formal guarantees}}\\ \midrule
    \rowcolor{lightgreen}
      COL (our method) & \cmark & \textbf{Causal structure} & \cmark \\
      CPO \cite{pmlr-v70-achiam17a}, P3O \cite{ijcai2022p520} & \xmark & System simulator & \xmark \\
      PPO-Lag \cite{ray2019benchmarking}, Safe SLACK \cite{hogewind2022safereinforcementlearningpixels} & \xmark & System simulator & \xmark \\      
      SafeDreamer \cite{ICLR2024_ece182f9} & \xmark & System simulator & \xmark \\            
      LAMBDA \cite{as2022constrainedpolicyoptimizationbayesian}, CCE \cite{NEURIPS2018_34ffeb35} & \xmark & System simulator & \xmark \\
      Saute RL \cite{sootla2022sauterlsurelysafe} & \xmark & System simulator & \cmark \\
      \cite{7798979,7330913,7039601,10.5555/3524938.3525846}, Shielding \cite{10.5555/3504035.3504361} & \cmark & Partially known dynamics & \cmark \\
      \cite{pmlr-v37-sui15,Fiedler2024OnSI,10.1109/ICRA48506.2021.9560738} & \cmark & Subset of safe region & \cmark \\
      \nocite{altman-constrainedMDP,9483029,ames_control_2019,9516971,zhou1998essentials,bertsekas2020constrainedmultiagentrolloutmultidimensional,Baier2008,larsen1997uppaal,belta2017formal,10.5555/1717907,belta2019formal,pmlr-v205-liu23e}[\citenum{altman-constrainedMDP}]--[\citenum{pmlr-v205-liu23e}] & \cmark & Known dynamics & \cmark \\                        
    \bottomrule\\
  \end{tabular}}
  \caption{Comparison of our method with related approaches for learning and validating the safety of dynamical systems.}\label{tab:related_work}
\end{table}

Our work is most closely related to the latter class of methods, as we similarly aim to learn the safe region of an operational system with unknown dynamics. However, our approach departs from prior work in two important ways. First, most existing methods are designed for physical or simulated control systems, whereas our focus is on cloud RANs. This distinction introduces new challenges because RANs exhibit properties that make them difficult to model with standard control-system assumptions, such as partial observability (due to limited monitoring coverage), non-stationarity (as workloads evolve), and discrete-time behavior (due to system metrics being sampled at fixed intervals). Second, whereas prior methods generally assume partial knowledge of the system dynamics or access to an initial safe region, we instead assume knowledge of the system’s \textit{causal structure}. We argue that this structure provides a more practical prior for learning safe regions in cloud RANs, which are engineered with well-defined control interfaces and causal structure even though their precise operational dynamics are unknown.

Lastly, we note that a growing body of research applies causal modeling to various management problems that arise in the operation of networked systems, such as failure recovery in cloud infrastructures \cite{10.1145/3698038.3698534}, performance optimization in IT systems \cite{hammar2025onlineidentificationsystemsactive}, causal discovery of microservice deployments \cite{NEURIPS2022_c9fcd02e,Wang_Rios_Jha_Shanmugam_Bagehorn_Yang_Filepp_Abe_Shwartz_2024,272110,10647072}, incident response planning in enterprise networks \cite{hammar2024optimaldefenderstrategiescage2,kim_phd_thesis}, and root-cause analysis in computing environments \cite{7563819}. While these works are related to our paper in terms of the general approach, they do not address the problem that we consider in this paper, namely the (online) learning of a cloud RAN's safe operating region, as detailed below.

\section{Use Case: Identifying the Safe Region \\of a Cloud Radio Access Network}\label{ex:cran}
We consider the problem of identifying the \textit{safe operating region} of a cloud radio access network (RAN), i.e., the set of \textit{control inputs} (e.g., resource allocations) for which the RAN's \textit{specification} is satisfied. This specification captures the service requirements that the RAN must meet, which are expressed as constraints over variables that can be measured from the RAN. As an example, the specification may require that the RAN maintain throughput above a threshold or that resource utilization remains within acceptable bounds. We say that a control input is \textit{safe} if it does not cause the RAN to violate the specification. By characterizing the region of such control inputs, we can safely explore alternative controls to optimize management objectives and adapt the RAN to changes.

We consider that the RAN is in operation and that its architecture and causal structure are known. However, the causal effects that define the safe region are assumed \textit{unknown}. For this reason, the safe region must be learned from \textit{system observations} (e.g., performance metrics), which can be obtained either by monitoring the RAN under its current configuration or by intervening on the RAN to change the configuration. During such an intervention, we set one or more control inputs to new values and validate the updated RAN behavior against the specification. For example, an intervention may involve updating the RAN's policy for traffic steering and validating whether this change keeps the spectral efficiency above a threshold \cite{10024837}. When selecting these interventions, the goal is to identify as much of the safe region as possible while keeping the risk of violating the specification low.
\begin{remark}
While we frame the task of learning the safe region as an online problem that involves an operational RAN rather than an offline problem that involves a fixed dataset of system observations, we do not require that the RAN be deployed in production. For example, the RAN may be deployed in a testbed or a digital twin where interventions can be applied without incurring operational cost and impacting live traffic.
\end{remark}

\begin{figure*}
  \centering
  \scalebox{1.25}{
   \includegraphics{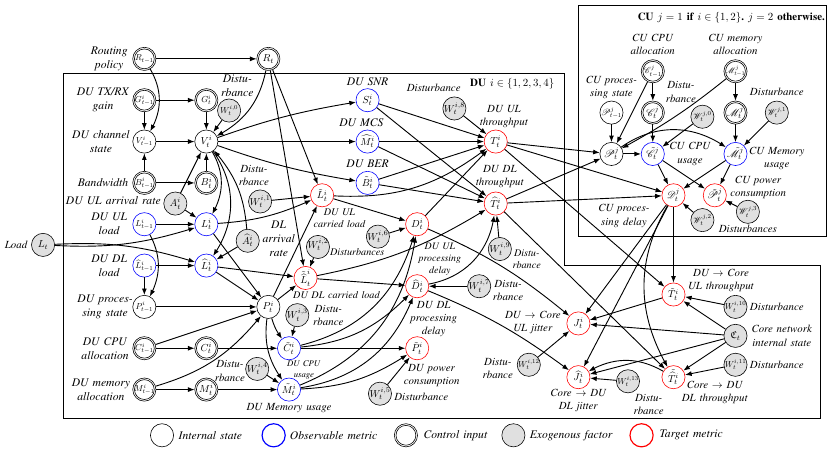}    
  }
\caption{Causal (summary) graph that represents the causal structure of the cloud radio access network in Fig.~\ref{fig:target_system}. For brevity, we use plate notation to represent the structure of sets of variables \cite{plate_notation}, i.e., the causal structure in the left box applies to the four distributed units (DUs) and the causal structure in the right box applies to the two central units (CUs). In total, the graph has $220$ nodes (system variables) and $367$ directed edges (causal dependencies). UL is an acronym for uplink and DL is an acronym for downlink. The SNR refers to the signal-to-noise ratio measured on the uplink of a DU. The MCS represents the average modulation and coding scheme across the uplink and downlink of a DU. Similarly, the BER represents the block error rate across the uplink and downlink of a DU. TX is a shorthand for transmitter and RX is a shorthand for receiver. Jitter is the variation in the delay (latency) of data packets.}
  \label{fig:causal_structure_5g}
\end{figure*}
\subsection*{Target System: 5G Cloud RAN}
We use the cloud RAN illustrated in Fig.~\ref{fig:target_system} as the target system under study in this paper. This RAN comprises two virtual baseband units (vBBUs), each of which runs on a shared cloud and implements a split baseband stack with multiple distributed units (DUs) connected to centralized units (CUs). These components realize four gNBs (5G base stations), each of which consists of a radio unit (RU), a DU, and a CU. Each RU is connected to a DU through the fronthaul and communicates with the user equipment (UE) over the 1800 MHz band. The DUs process physical-layer functions and forward traffic from the RUs over the midhaul to the CUs, which handle higher-layer processing and aggregate traffic to the core network via the backhaul. Finally, the core manages session data and provides connectivity to the data network. 

In addition to the core network, the DUs and CUs interface with a near-real-time RAN intelligent controller (near-RT RIC) deployed in the edge cloud, which in turn interfaces with a non-real-time RIC hosted in an external service management and orchestration (SMO) platform. These controllers provide dynamic management and optimization of various network functions, such as load balancing and scheduling coordination. The near-RT RIC operates on time scales of 10ms to 1s, while the non-RT RIC works on time scales of seconds to minutes.

\section{System Model}
We model the RAN in Fig.~\ref{fig:target_system} formally as a discrete-time dynamical system that evolves through unknown dynamics over time steps $(t=0,1,\hdots)$, where the duration of each time step is determined by the RAN’s monitoring interval (e.g., $15$ seconds). At each step, the system comprises internal states $\mathbf{X}_t \in X$, observable metrics $\mathbf{O}_t \in O$, control inputs $\mathbf{U}_t \in U$, target metrics $\mathbf{Y}_t \in Y$, and exogenous factors $\mathbf{W}_t \in W$. 

The causal structure among these variables is encoded in the \textit{causal graph} in Fig.~\ref{fig:causal_structure_5g}, where nodes correspond to variables and edges indicate causal influence. This graph shows the causal structure for a single time step of the RAN, i.e., it shows the system variables at time $t$ and their causal dependencies, without depicting dependencies of variables at other time steps of the RAN's evolution. We have derived the graph from knowledge of the RAN’s configuration and validated it against data from our 5G testbed; see \S\ref{experimental_evaluation} for details.

In total, the graph includes $220$ variables and $367$ causal dependencies. We summarize the key variable types below.
\begin{itemize}
\item \textbf{\textit{Exogenous factors} $\mathbf{W}_t$} include external disturbances and network traffic characteristics, such as the packet arrival rate on the uplink of $\text{DU}_i$, which we denote by $A^i_t$.
\item \textbf{\textit{Internal states} $\mathbf{X}_t$} represent internal (hidden) system properties, such as the processing state $\mathscr{P}^{j}_t$ and the channel state $V^i_t$, which describe the processing conditions of $\text{CU}_j$ and the radio conditions of $\text{DU}_i$, respectively.
\item \textbf{\textit{Observable metrics} $\mathbf{O}_t$} are measurements available through monitoring systems, such as the measured load on the uplink of $\text{DU}_i$ (denoted by $L^i_t$), the CPU usage percentage of $\text{CU}_j$ (denoted by $\tilde{\mathscr{C}}^j_t$), and the measured (uplink) signal-to-noise ratio of $\text{DU}_i$ (denoted by $S^i_t$).
\item \textbf{\textit{Control inputs} $\mathbf{U}_t$} correspond to the configurable variables in the RAN. They include the CPU allocations (e.g., $C^i_t$), the memory allocations (e.g., $M^i_t$), and the bandwidth allocations (e.g., $B^i_t$), among others.
\item \textbf{\textit{Target metrics} $\mathbf{Y}_t$} represent key performance indicators, such as the processing delay of $\text{CU}_j$ (denoted by $\mathscr{P}^j_t$) and the downlink throughput of $\text{DU}_i$ (denoted by $\Tilde{\widehat{T}^i}_t$).
\end{itemize}
During the operation of the RAN, the target metrics can be measured through the RAN’s monitoring infrastructure. These measurements allow to validate the RAN's behavior against its \textit{specification}, i.e., the operational constraints that the RAN should satisfy. We express these constraints formally through a Boolean function of the target metrics, as defined below.
\begin{definition}[Specification]\label{def:specification}
Given a RAN with target metrics $\mathbf{Y}_t \in Y$, we define the specification $\varphi$ as a Boolean function applied to a system trajectory of length $H+1$, i.e.,
\begin{align*}
\varphi: Y^{H+1} \mapsto \{0,1\},
\end{align*}
where $H$ is the \emph{evaluation horizon}.
\end{definition}
This definition captures both constraints that depend only on the current state of the RAN and constraints that depend on how the RAN evolves over a time horizon $H$. For example, a specification with horizon $H=1$ could check that a control does not cause the jitter of the packet arrival times to exceed a threshold at the next time step. Similarly, a specification with $H=100$ could check that a performance metric stays within bounds as the RAN evolves over the next 100 time steps.

To ensure that the RAN meets the specification, an operator can intervene and modify the values of the control inputs $\mathbf{U}_t$. We model such \textit{interventions} with the do-operator as $\mathrm{do}(\mathbf{U}'_t = \mathbf{u}'_t)$, which represents the operation of changing the variables in the set $\mathbf{U}'_t \subseteq \mathbf{U}_t$ to the values in the vector $\mathbf{u}'_t$ \cite{pearl2000causality}. The evaluation horizon $H$ determines how long the RAN must be monitored after such an intervention to decide whether it operates according to the specification. Specifically, the \textit{causal effect} on the specification when assigning the values in the vector $\mathbf{u}'_t$ to the variables in the set $\mathbf{U}'_{t}$ is given by
\begin{align}
p_{\mathbf{U}'_{t}}(\mathbf{u}'_t) &= P\big(\varphi(\mathbf{Y}_{t},\hdots,\mathbf{Y}_{t+H})=1 \mid \mathrm{do}(\mathbf{U}'_t = \mathbf{u}'_t)\big).\label{eq:intervention_def}
\end{align}
We define the subset of interventions for which the specification $\varphi$ is satisfied with high probability to be the \textit{safe region} of the control space, as expressed by the following definition.
\begin{definition}[Safe region]\label{def:safe_region}
We define the \textit{safe region} at time $t$ for the control inputs $\mathbf{U}'_t\subseteq \mathbf{U}_t$ as the set
\begin{align*}
\mathcal{S}_{\mathbf{U}'_t} &= \bigg\{ \mathbf{u}'_t \Bigm| \mathbf{u}'_t \in U', \text{ }p_{\mathbf{U}'_t}(\mathbf{u}'_t) \geq \delta\bigg\},
\end{align*}
where $p_{\mathbf{U}'_t}(\mathbf{u}'_t)$ is the probability of meeting the specification $\varphi$ [cf.~Def.~\ref{def:specification}] given the intervention $\mathrm{do}(\mathbf{U}'_t = \mathbf{u}'_t)$ [cf.~\eqref{eq:intervention_def}], $\delta\in (0,1)$ is the required safety level, and $U' \subseteq U$ is the control space of the variables in the set $\mathbf{U}'_t$.
\end{definition}
\section{Problem: Online Learning of the Safe Region}
Given the above definitions, we consider the problem of learning the safe operating region from system observations. We formulate this problem as an (online) active learning problem that involves two interconnected tasks: (\textit{i}) estimating the safe region based on the available observations; and (\textit{ii}) selecting interventions for collecting new observations. These tasks are carried out in an iterative process, where each new intervention informs the next estimate of the safe region.

Formally, we model active learning of the safe region as a discrete-time process in which interventions and estimations occur at time steps $t_0, t_1,\hdots, t_{K}$. At each step $t_k$, one intervention is performed, after which we monitor the system for $H$ steps and validate whether it satisfies the specification; cf.~Def.~\ref{def:specification}. We then use the resulting observations to update our estimate of the safe region and select the next intervention. When designing this iterative process, our goal is to quickly learn the safe region while meeting the system specification with a given confidence $\alpha$ and keeping intervention costs below a specified level $C$. We formalize this objective as
\begin{subequations}\label{eq:problem_statement}
\begin{align}
  &\maximize_{K,\pi_{\mathbf{U}'_{t_k}},\phi_{\mathbf{U}'_{t_k}}} \quad\lambda(\hat{\mathcal{S}}_{\mathbf{U}_{t_K}'}), \\
  &\text{ }\text{$\mathrm{subject}$ $\mathrm{to}$} \quad\text{ } \hat{\mathcal{S}}_{\mathbf{U}'_{t_k}} = \phi_{\mathbf{U}'_{t_k}}(\mathcal{D}_{t_k}), \label{eq:estimator}\\
  &\quad\quad\quad\quad\quad\text{ }\text{ }\text{ } \mathrm{do}(\mathbf{U}'_{t_k}=\mathbf{u}'_{t_k})=\pi_{\mathbf{U}'_{t_k}}(\hat{\mathcal{S}}_{\mathbf{U}'_{t_k}}), \label{eq:policy_def}\\
  &\quad\quad\quad\quad\quad\text{ }\text{ }\text{ }P(\hat{\mathcal{S}}_{\mathbf{U}'_{t_k}} \subseteq \mathcal{S}_{\mathbf{U}'_{t_k}}) \geq \alpha,\label{eq:safety_constraint}\\
  & \quad\quad\quad\quad\quad\text{ }\text{ }\text{ }\sum_{k=0}^Kc(\mathbf{u}'_{t_k}) \leq C,\label{eq:cost_budget}
\end{align}
\end{subequations}
for all intervention steps $t_k$ and sets of control inputs $\mathbf{U}'_{t_k}$.

Here $\mathcal{D}_{t_k}$ is the dataset of system observations at time $t_k$, $\phi_{\mathbf{U}'_{t_k}}$ is an estimator of the safe region, $\pi_{\mathbf{U}'_{t_k}}$ is an intervention policy, and $\lambda$ is a function that measures the size of the estimated region. For instance, if the estimated region is a compact subset of a Euclidean space, then $\lambda$ can be the Lebesgue measure. In discrete settings, $\lambda$ may instead be a counting measure. Equation \eqref{eq:safety_constraint} expresses that the estimated region should be a subset of the safe region with probability at least $\alpha$, where $\alpha\in (0,1)$ is the desired \textit{confidence level}. This confidence is linked to the safety level $\delta$ in Def.~\ref{def:safe_region} in the sense that increasing $\delta$ lowers the achievable $\alpha$ in practice. Finally, \eqref{eq:cost_budget} defines the cost constraint, where $c$ is a configurable function that encodes the cost of interventions on the RAN.

The optimization problem in \eqref{eq:problem_statement} is generally intractable as the objective is non-convex and the safe region may have a disconnected structure. Moreover, satisfying constraint \eqref{eq:safety_constraint} may require infinitely many interventions. As a result, approximation methods are required for learning the safe region in practice. We present such a method in the next section.

\section{Our Method: \textbf{C}ausal \textbf{O}nline \textbf{L}earning (COL)}
\begin{figure}[H]
  \centering    
  \scalebox{1}{
   \includegraphics{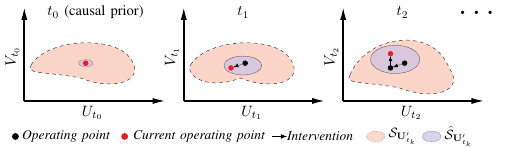}
  }
\caption{Schematic of our method for identifying the safe region $\mathcal{S}_{\mathbf{U}'_{t_k}}$ of control inputs: $\mathbf{U}'_{t_k} = \{U_{t_k},V_{t_k}\}$. We first observe the RAN for $t_0$ time steps to infer an initial safe region $\hat{\mathcal{S}}_{\mathbf{U}'_{t_0}}$ via causal inference, after which we gradually expand the estimated region through a sequence of interventions.}
\label{fig:safe_region_1}
\end{figure}

We address problem \eqref{eq:problem_statement} by developing an iterative method for learning the safe region of a cloud RAN during operation. The method, which we call \textbf{C}ausal \textbf{O}nline \textbf{L}earning (COL), can be divided into two phases, both of which are executed online. In the first phase, we observe the RAN at the current operating point to infer an initial safe region via causal inference and Gaussian process (GP) regression. In the second phase, we gradually expand this region through active interventions and Bayesian learning; see Fig.~\ref{fig:safe_region_1}. 

\subsection{Initializing the Safe Region through Causal Inference}
Before intervening on the RAN, we need to identify an initial safe region from which exploration can begin without risking violation of the specification. To accomplish this, the first step of our method is to observe the RAN as it operates under its current configuration and estimate the causal effect of control inputs on the specification $\varphi$ via \textit{causal inference}.

Following the mathematical theory of causality \cite{pearl2000causality}, we say that an intervention is \textit{identifiable} if its causal effect on the specification can be inferred from system observations obtained through passive monitoring of the RAN. Formally,
\begin{definition}[Causal effect identifiability]\label{def:ID}
The causal effect of the intervention $\mathrm{do}(\mathbf{U}'_{t_k}=\mathbf{u}'_{t_k})$ on the specification $\varphi$ is identifiable from system observations if the probability 
\begin{align*}
p_{\mathbf{U}'_{t_k}}(\mathbf{u}'_{t_k}) = P\big(\varphi(\mathbf{Y}_{t_k},\hdots,\mathbf{Y}_{t_k+H})=1 \mid \mathrm{do}(\mathbf{U}'_{t_k} = \mathbf{u}'_{t_k})\big)
\end{align*}
is \emph{uniquely} computable from the observational probability distribution $P((\mathbf{O}_l,\mathbf{U}_l,\mathbf{Y}_l)^{l=t_k+H}_{l=t_k})$ and the causal graph.
\end{definition}
The causal effect of an identifiable intervention can be derived via do-calculus, as defined by Pearl \cite[Thm. 3.4.1]{pearl2000causality}. For instance, given the causal graph in Fig.~\ref{fig:causal_structure_5g}, the effect on the power consumption $\Tilde{P}^i_{t}$ caused by changing the CPU allocation $C^i_t$ of $\text{DU}_i$ can be derived via do-calculus as
\begin{align*}
P(\Tilde{P}^i_{t}|\mathrm{do}(C^i_t=c^i_t)) 
&\numeq{a} \sum_{c^{i}_{t-1}} P(\Tilde{P}^i_{t}|\mathrm{do}(C^i_t=c^i_t), C^i_{t-1}=c^{i}_{t-1}) \\
&\quad\quad\quad \cdot P(C^i_{t-1}=c^{i}_{t-1} \mid \mathrm{do}(C^i_t=c^i_t)) \\
&\numeq{b} \sum_{c^{i}_{t-1}} P(\Tilde{P}^i_{t} \mid C^i_t=c^i_t, C^i_{t-1}=c^{i}_{t-1}) \\
&\quad\quad\quad \cdot P(C^i_{t-1}=c^{i}_{t-1} \mid \mathrm{do}(C^i_t=c^i_t)) \\
&\numeq{c} \sum_{c^{i}_{t-1}} P(\Tilde{P}^i_{t} \mid C^i_t=c^i_t, C^i_{t-1}=c^{i}_{t-1}) \\
&\quad\quad\quad \cdot P(C^i_{t-1}=c^{i}_{t-1}),
\end{align*}
where (a) follows from the law of total probability, (b) follows from Rule 2 of do-calculus, and (c) follows from Rule 3. Since this expression consists of regular probability distributions, it can be estimated without having to intervene on the RAN.

We leverage such inference in our method to learn the causal effect of all identifiable interventions based on the available system observations. Specifically, in the first phase of COL, we passively observe the RAN for $t_0$ time steps and collect an initial dataset $\mathcal{D}_{t_0}$ of system observations, where $t_0 > 0$ is a configurable parameter. We then use this dataset in combination with the causal graph to infer the effects of all identifiable interventions, which provides us with an initial estimate of the safe operating region. In particular, let $\mu_{\mathbf{U}'_{t_0}}(\mathbf{u}'_{t_0})$ denote the inferred causal effect of the intervention $\mathrm{do}(\mathbf{U}'_{t_0}=\mathbf{u}'_{t_0})$ on the specification $\varphi$. Given this notation, we define the initial safe region for the control inputs $\mathbf{U}'_{t_0}$ as
\begin{align}\label{eq:initial_region}
  \hat{\mathcal{S}}_{\mathbf{U}'_{t_0}} &= \bigg\{ \mathbf{u}'_{t_0} \Bigm| \mathbf{u}'_{t_0} \in U', \text{ }\mu_{\mathbf{U}'_{t_0}}(\mathbf{u}'_{t_0}) \geq \delta, \\
  &\quad \quad\mathrm{do}(\mathbf{U}'_{t_0}=\mathbf{u}'_{t_0}) \text{ is identifiable; cf.~Def.~\ref{def:ID}.}\bigg\}. \nonumber
\end{align}
\subsection{Bayesian Learning of the Safe Region}
Having identified an initial estimate of the safe region using the inference described above, the next task is to gradually expand this estimate by performing interventions in areas of the control space where the RAN is likely but not yet certain to satisfy the specification. To perform this exploration efficiently while ensuring that the specification remains satisfied with high probability, we require a model of the probability that each intervention $\mathrm{do}(\mathbf{U}'_{t_k}=\mathbf{u}'_{t_k})$ satisfies the specification.

In our method, we construct such a model using a Gaussian process (GP) $\hat{f}_{\mathbf{U}'_{t_k}} \sim \mathcal{GP}(m_{\mathbf{U}'_{t_k}}, k_{\mathbf{U}'_{t_k}})$, where $m_{\mathbf{U}'_{t_k}}$ is the mean function and $k_{\mathbf{U}'_{t_k}}$ is the covariance function. This GP defines a prior distribution over functions $\hat{f}_{\mathbf{U}'_{t_k}}$, where $\hat{f}_{\mathbf{U}'_{t_k}}(\mathbf{u}'_{t_k})$ is an estimate of the causal effect $p_{\mathbf{U}'_{t_{k}}}(\mathbf{u}'_{t_k})$, whose uncertainty is quantified by the variance $k_{\mathbf{U}'_{t_k}}(\mathbf{u}', \mathbf{u}')$. Similarly, the covariance $k_{\mathbf{U}'_{t_k}}(\mathbf{u}', \mathbf{u}'')$ encodes the correlation between the estimated causal effects of interventions with different controls $\mathbf{u}'$ and $\mathbf{u}''$. We initialize these functions as
\begin{equation}\label{gp_init}
\begin{aligned}
m_{\mathbf{U}'_{t_{k}}}(\mathbf{u}') &\!\!=\!\! \mu_{\mathbf{U}'_{t_k}}(\mathbf{u}'),\\
k_{\mathbf{U}'_{t_k}}(\mathbf{u}', \mathbf{u}'') &\!\!=\!\! \sigma_{\mathbf{U}'_{t_k}}(\mathbf{u}')\sigma_{\mathbf{U}'_{t_k}}(\mathbf{u}'')\exp\left(-\frac{\norm{\mathbf{u}'\!\!-\!\!\mathbf{u}''}^2}{2}\right),
\end{aligned}
\end{equation}
where $\mu_{\mathbf{U}'_{t_0}}$ and $\sigma_{\mathbf{U}'_{t_k}}$ are the mean and standard deviation functions inferred in the first phase of COL; cf.~\eqref{eq:initial_region}. Hence, the inference phase provides a \textit{causal prior} for the GP.

Once the GP is initialized, we update it through Bayesian learning from system observations. At each stage of this learning process, we select an intervention $\mathrm{do}(\mathbf{U}_{t_k}'=\mathbf{u}_{t_k})$ and apply it to the RAN, after which we monitor the RAN for $H$ time steps and validate whether it satisfies the specification. We then add the resulting system observations to the dataset $\mathcal{D}_{t_{k+1}}$ and use it to update the GP prior [cf.~\eqref{gp_init}] via Bayes' rule.\footnote{See Rasmussen and Williams for formulas of these updates \cite[Def. 2.1]{Rasmussen2006Gaussian}.} We denote the resulting (posterior) GP as
\begin{align}
(\hat{f}_{\mathbf{U}'_{t_{k+1}}} \mid \mathcal{D}_{t_{k+1}}) \sim \mathcal{GP}(m_{\mathbf{U}'_{t_{k+1}}|\mathcal{D}_{t_{k+1}}}, k_{\mathbf{U}'_{t_{k+1}}|\mathcal{D}_{t_{k+1}}}),\label{eq:gp_bayes}
\end{align}
where $m_{\mathbf{U}'_{t_{k+1}}|\mathcal{D}_{t_{k+1}}}$ and $k_{\mathbf{U}'_{t_{k+1}}|\mathcal{D}_{t_{k+1}}}$ denote the posterior mean and covariance functions, respectively. Given suitable regularity conditions, the mean of this posterior converges to the true probability that each intervention satisfies the specification, as formally stated in the following proposition.

\begin{proposition}\label{prop:consistency}
Assume a) that samples in the dataset $\mathcal{D}_{t_k}$ are independent and identically distributed (i.i.d.); and b) that the difference between the causal effect $p_{\mathbf{U}'_{t_k}}(\mathbf{u}'_{t_k})$ and the (prior) mean function $m_{\mathbf{U}'_{t_k}}$ lies in the reproducing kernel Hilbert space (RKHS) of the covariance function $k_{\mathbf{U}'_{t_k}}$ for all control settings $\mathbf{u}_{t_k}' \in U'$. If each $\mathbf{u}_{t_k}'$ is sampled infinitely often with independent zero-mean and bounded Gaussian noise $\epsilon$, then
\begin{align*}
\lim_{|\mathcal{D}_{t}|\to \infty}E\left\{\left(\hat{f}_{\mathbf{U}'_{t_k}}(\mathbf{u}_{t_k}')-p_{\mathbf{U}'_{t_k}}(\mathbf{u}'_{t_k})\right)^2\right\} = 0,
\end{align*}
for all $\mathbf{u}'_{t_k} \in U'$, where $\hat{f}_{\mathbf{U}'_{t_k}} = E_{f \sim (\hat{f}_{\mathbf{U}'_{t_k}} \mid \mathcal{D}_{t_k})}\{f\}$ and the expectation is with respect to the sampling noise.
\end{proposition}
Proposition~\ref{prop:consistency} implies that the GP can recover the true probability that each intervention satisfies the specification in the limit of infinite data. This is a well-known result in GP theory, see e.g., \cite[Thm. 3]{JMLR:v22:21-0853} for a detailed analysis and proof. (Note that this result only applies when the RAN is stationary so that the causal effects remain unchanged over time.)

Beyond this asymptotic consistency, the GP also provides confidence intervals for its predictions, i.e., it allows us to quantify the uncertainty of the predicted causal effects. By using these confidence intervals, we can construct a principled estimate of the safe region that satisfies constraint \eqref{eq:safety_constraint} under certain conditions, as formalized in the next proposition.
\begin{proposition}\label{prop:safe_confidence}
Assume conditions a)-b) of Prop.~\ref{prop:consistency} and suitable regularity conditions (see Appendix~\ref{app:gp_bound}) hold. If
\begin{align}
\hat{\mathcal{S}}_{\mathbf{U}'_{t_k}} = \bigg\{\mathbf{u}' \Bigm| \mathbf{u}' \in U', m_{\mathbf{U}'_{t_k}|\mathcal{D}_{t_k}}(\mathbf{u}')-\kappa_{t_k}(\mathbf{u}') \geq \delta \bigg\},\label{eq:region_estimate}
\end{align}
where $\kappa_{t_k}(\mathbf{u}')$ is a suitable constant (see Appendix~\ref{app:gp_bound}), then
\begin{align*}
P\big(\hat{\mathcal{S}}_{\mathbf{U}'_{t_k}} \subseteq \mathcal{S}_{\mathbf{U}'_{t_k}}\big) \geq \alpha, && \text{for all }t_k\geq 0,
\end{align*}
where $\alpha \in (0,1)$ is the chosen confidence level; cf.~\eqref{eq:safety_constraint}.
\end{proposition}
Due to the technical nature of this proposition, we defer the proof to Appendix~\ref{app:gp_bound}. Proposition \ref{prop:safe_confidence} provides a theoretical justification for using \eqref{eq:region_estimate} as an estimate of the safe region given the available data. In the following, we describe our intervention policy for collecting this data in an efficient way.
   
\subsection{Selecting Interventions through Active Learning}
Given the estimator \eqref{eq:region_estimate} of the safe region, the next challenge is to select interventions to efficiently collect data for improving the estimate while satisfying safety constraint \eqref{eq:safety_constraint}. To formalize this trade-off between safety, exploration, and cost, we use an intervention policy that prioritizes interventions within the estimated safe region with the highest expected information gain per unit cost. Specifically, we select interventions using a policy that is defined by the decision rule 
\begin{align}
\pi_{\mathbf{U}'_{t_k}}(\hat{\mathcal{S}}_{\mathbf{U}'_{t_k}}) \in \argmax_{\mathbf{u} \in \hat{\mathcal{S}}_{\mathbf{U}'_{t_k}}}\left\{\frac{\sqrt[]{k_{\mathbf{U}'_{t_k}|\mathcal{D}_{t_k}}(\mathbf{u}, \mathbf{u})}}{c(\mathbf{u})}\right\},\label{eq:policy}
\end{align}
where $c$ is the intervention-cost function [cf.~\eqref{eq:cost_budget}] and $k_{\mathbf{U}'_{k_t}|\mathcal{D}_{k_t}}$ is the covariance function of the GP, which captures the expected information gain of the intervention \cite{gp_ucb}.

This intervention policy is inspired by acquisition functions used in Bayesian optimization, see e.g., \cite{pmlr-v37-sui15}. It strikes a balance between optimality and computational tractability in the sense that it selects interventions according to a principled criterion, but does so in a greedy manner by optimizing only the immediate expected gain without planning multiple interventions ahead. In principle, an optimal intervention policy can be computed via dynamic programming; see Bertsekas \cite{bertsekas2022rolloutalgorithmsapproximatedynamic}. However, this computation is intractable in practice due to the control space $U$ being high-dimensional and continuous.
\subsection{Summary of Our Method for Identifying the Safe Region}
In summary, our method for identifying the safe region of a cloud RAN consists of two main phases. First, we use observations of the RAN in its current state to estimate an initial safe region via causal inference. Second, we expand the estimated region through Bayesian learning with Gaussian processes (GPs) based on observations collected through a sequence of controlled interventions, which are chosen to maximize the expected information gain while being safe. We refer to this method as \textbf{C}ausal \textbf{O}nline \textbf{L}earning (COL). The complete procedure of COL is summarized in Alg.~\ref{alg:our_method}.  
\begin{algorithm}
\footnotesize
\DontPrintSemicolon
\caption{\textbf{C}ausal \textbf{O}nline \textbf{L}earning.}\label{alg:our_method}

\nonl \textbf{Input:} Monitoring steps $t_0$, time horizon $H$, intervention set $\mathbf{U}'$,\\
\nonl $\quad\quad\quad$ cost function $c$, cost budget $C$, and confidence level $\alpha$.\;
\nonl \textbf{Output:} An estimated safe region $\hat{\mathcal{S}}_{\mathbf{U}'_{t_k}}$.\;
Monitor the RAN for $t_0$ time steps and collect observations.\;
Initialize the safe region via causal inference according to \eqref{eq:initial_region}.\;
Initialize the GP via causal inference according to \eqref{gp_init}.\;
Initialize $t_k=t_0$, $k=0$, and $\tilde{c} =0$.\;
\While{$\tilde{c} \leq C$}{
Select the intervention $\mathrm{do}(\mathbf{U}'_{t_k}=\mathbf{u}_{t_k})$ through \eqref{eq:policy}.\;
\If{$\tilde{c} + c(\mathbf{u}_{t_k}) \leq C$}{
Carry out intervention $\mathrm{do}(\mathbf{U}'_{t_k}=\mathbf{u}_{t_k})$ on the RAN.\;
Collect a trajectory of length $H+1$ and update $\mathcal{D}_{t_{k}+H}$.\;
Update $t_{k+1} = t_k\!\!+H\!\!+\!\!1$, $k=k+1$, and $\tilde{c} = \tilde{c} + c(\mathbf{u}_{t_k})$.\;
Update the GP via Bayes rule; cf.~\eqref{eq:gp_bayes}.\;
Update the estimated safe region via \eqref{eq:region_estimate}.\;
}
}
Return the estimated safe region $\hat{\mathcal{S}}_{\mathbf{U}'_{t_k}}$.\;
\normalsize
\end{algorithm}
\begin{table*}
  \centering
  \scalebox{0.77}{
    \begin{tabular}{ccccc} \toprule
\rowcolor{lightgray}
      {\textit{Method}} & {\textit{Number of unsafe interventions to converge}} & {\textit{Output}} & {\textit{Size of estimated safe region, i.e., $\lambda(\widehat{\mathcal{S}}_{\mathbf{U}'_t})$}} & {\textit{Formal guarantees}} \\ \midrule
      \multicolumn{5}{c}{\underline{\textbf{\textit{Simulation Scenario 1: Steady state system with a fixed safe region.}}}} \\      
    \rowcolor{lightgreen}
      COL [cf.~Alg.~\ref{alg:our_method}] & $\bm{6.10 \pm 0.30}$ & Safe region $\widehat{\mathcal{S}}_{\mathbf{U}'_t}$  & $\bm{0.48 \pm 0.05}$ & The estimated region is safe with probability $\alpha=0.8$\\
      CPO \cite{pmlr-v70-achiam17a} & $60342.33 \pm 81.12$ & Safe control policy & $0.01 \pm 0.01$ (inferred from the control policy) & None\\
      PPO-Lagrangian \cite{ray2019benchmarking} & $63271.76 \pm 104.77$ & Safe control policy & $0.01 \pm 0.01$ (inferred from the control policy) & None\\
      \multicolumn{5}{c}{$\quad$} \\             
      \multicolumn{5}{c}{\underline{\textbf{\textit{Simulation Scenario 2: Non-stationary system with a dynamic safe region.}}}} \\        \rowcolor{lightgreen}
      COL [cf.~Alg.~\ref{alg:our_method}] & $\bm{19.80 \pm 2.41}$ & Safe region $\widehat{\mathcal{S}}_{\mathbf{U}'_t}$  & $\bm{0.12 \pm 0.04}$ & The estimated region is safe with probability $\alpha=0.8$\\
      CPO \cite{pmlr-v70-achiam17a} & $23412.41 \pm 89.24$ & Safe control policy & $0.01 \pm 0.00$ (inferred from the control policy) & None\\
      PPO-Lagrangian \cite{ray2019benchmarking} & $64292.61 \pm 1292.92$ & Safe control policy & $0.01 \pm 0.00$ (inferred from the control policy) & None\\
    \bottomrule\\
  \end{tabular}}
  \caption{Simulation results of COL and safe reinforcement learning baselines when applied to the example system in Fig.~\ref{fig:example_system}. Numbers indicate the mean and standard deviation from $10$ evaluations with different random seeds.}\label{tab:simulation_safe_rl}
\end{table*}
\section{Illustrative Example}\label{sec:ill_example}
\begin{figure}[H]
  \centering
  \scalebox{0.95}{
   \includegraphics{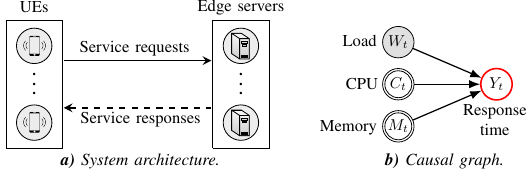}    
  }
  \caption{Example networked system [cf.~\textbf{a})] and its causal structure [cf.~\textbf{b})]. The system consists of edge servers that handle service requests from UEs.}
  \label{fig:example_system}
\end{figure}

Before applying COL to learn the safe region of the RAN in Fig.~\ref{fig:target_system}, we illustrate COL by applying it to a simulated system with edge servers that handle service requests from UEs. The system architecture is shown in Fig.~\ref{fig:example_system} and involves three main variables: the response time of a service request $Y_t \in \mathbb{R}_{> 0}$, the CPU allocation of the edge servers $C_t \in [0,1]$, and the memory allocation of the edge servers $M_t \in [0,1]$. 

Suppose that the response time must remain below $50$\,ms over a time horizon of length $H$. We can express this specification through signal temporal logic as
\begin{align}
\varphi(Y_t,\hdots,Y_{t+H}) = \square_{\tau \in [t,t+H]}Y_{\tau} < 50,\label{eq:example_specification}
\end{align}
where $\square_{\tau \in [t,t+H]}$ denotes the \textit{always} logic operator over the time interval $[t,t+H]$, which means that the predicate $Y_{\tau} < 50$ must hold for all times $\tau$ within the horizon $[t,t+H]$.

In addition to the controls (i.e., the CPU and memory allocations), the response time also depends on the load $W_t \in [0,1]$, which models the (normalized) number of service requests per second, as expressed by the causal graph in Fig.~\ref{fig:example_system}.

\begin{figure*}
  \centering      
  \scalebox{0.7}{
   \includegraphics{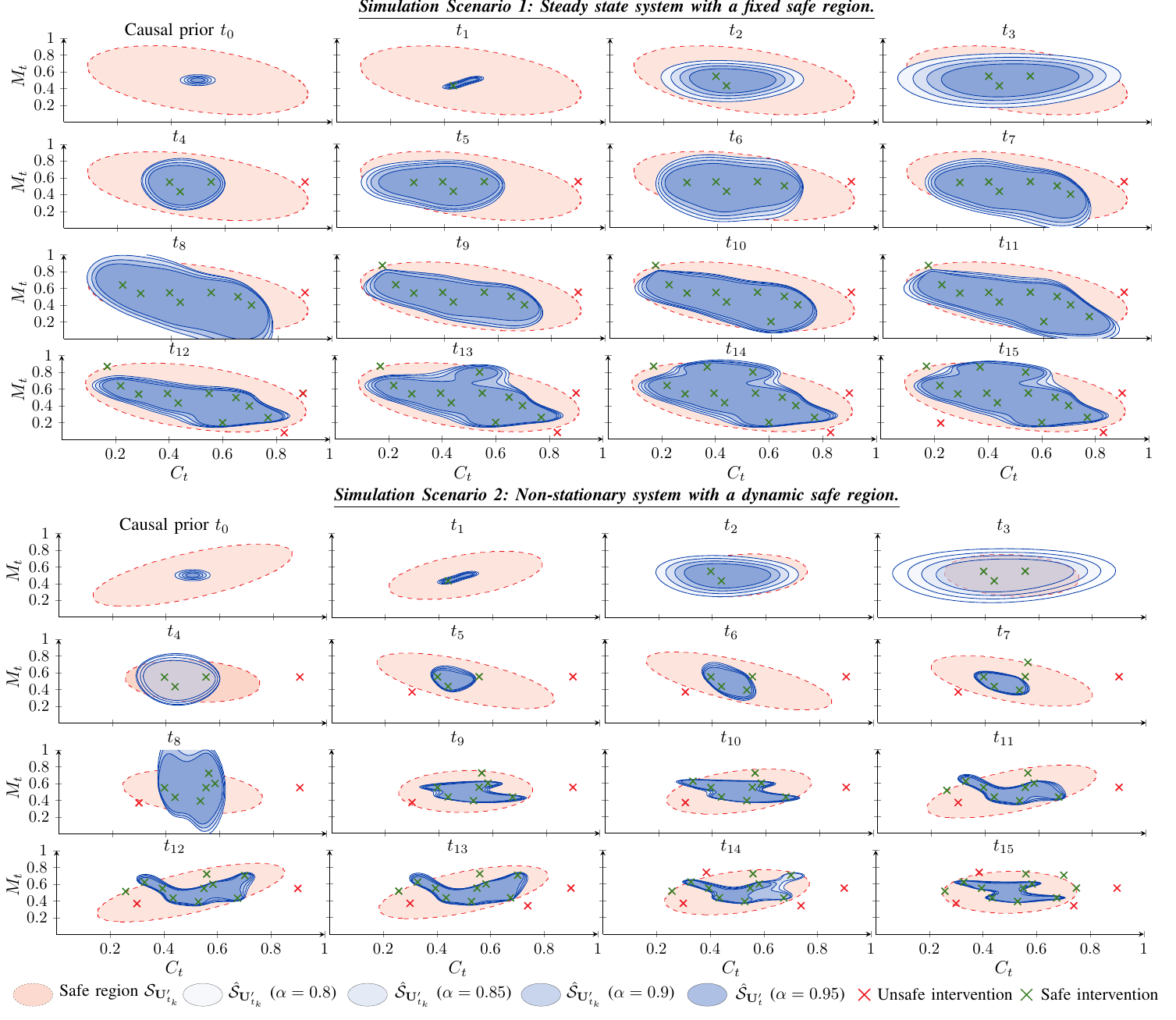}
  }
\caption{Evolution of the safe region $\hat{\mathcal{S}}_{\mathbf{U}'_{t_k}}$ estimated with COL [cf.~Alg.~\ref{alg:our_method}] for the simulated example system in Fig.~\ref{fig:example_system}. We instantiate COL with the control inputs $\mathbf{U}'_{t_k}=\{C_{t_k}, M_{t_k}\}$ and $t_0=10$ monitoring steps. The true safe region $\mathcal{S}_{\mathbf{U}'_{t_k}}$ is indicated in red color; the estimated region $\hat{\mathcal{S}}_{\mathbf{U}'_{t_k}}$ is indicated in blue color. The shaded blue regions indicate regions with confidence levels $\alpha=0.8$, $\alpha=0.85$, $\alpha=0.9$, and $\alpha=0.95$ (lighter to darker shades of blue).}\label{fig:example_learning}
\end{figure*}

\begin{figure*}
  \centering      
  \scalebox{0.63}{
   \includegraphics{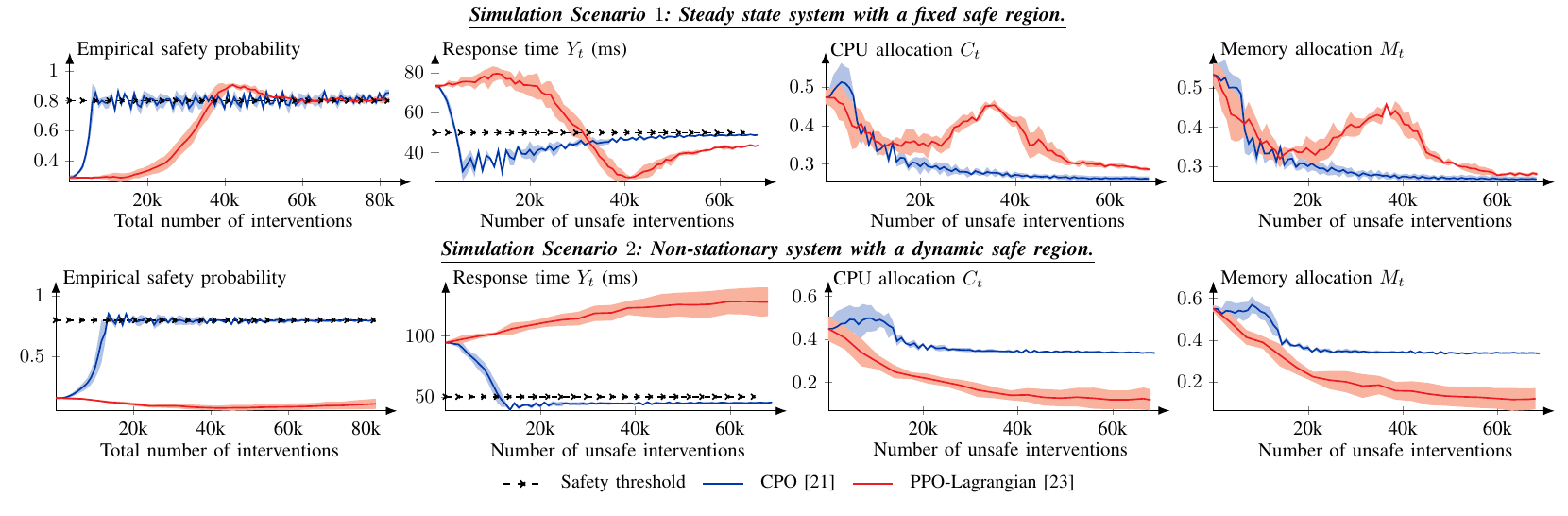}
  }
\caption{Learning curves of the safe reinforcement learning baselines when applied to the simulated example system in Fig.~\ref{fig:example_system}. Curves show the mean values based on $10$ evaluations; shaded regions indicate standard deviations. The x-axes indicate the number of interventions that violate specification \eqref{eq:example_specification}.}\label{fig:safe_rl_2}
\end{figure*}

\vspace{2mm}
\noindent\textit{\textbf{Example instantiation of COL.}}
We instantiate COL for the example system with evaluation horizon $H=1$, safety level $\delta=0.8$, confidence level $\alpha = 0.8$, $t_0=10$ monitoring steps, cost budget $C=20$, and intervention set $\mathbf{U}'_{t_k} = \{C_{t_k}, M_{t_k}\}$. It follows from the causal graph in Fig.~\ref{fig:example_system} that the causal effect on the response time of such an intervention is given by
\begin{align*}
P(Y_{t_k} \mid \mathrm{do}(\mathbf{U}'_{t_k} = \mathbf{u}'_{t_k})) &= P(Y_{t_k} \mid \mathbf{U}'_{t_k} = \mathbf{u}'_{t_k}), && \text{for all }t_k.
\end{align*}
We define the cost of the intervention as
\begin{align*}
c(\mathbf{u}'_{t_k}) = \left( (1+C_{t_k} - 0.5)^2 + (1+M_{t_k} - 0.5)^2 \right),
\end{align*}
i.e., the cost of the intervention increases as the values assigned to the CPU $C_{t_k}$ and memory $M_{t_k}$ increase above $0.5$.

\vspace{2mm}
\noindent\textit{\textbf{Simulation scenarios.}} We consider two simulation scenarios: one focuses on identifying the safe region during steady-state operation, and the other on tracking the safe region as it changes over time. In both scenarios, the dynamics of the CPU allocation $C_t$ and the memory allocation $M_t$ are defined as
\begin{align*}
C_t,M_t &\sim \mathrm{Beta}(0.5, 0.5), && \text{for all }t.
\end{align*}
The scenarios differ in the load distribution $P(W_t)$ and the dynamics of the response time $Y_t$. In Scenario $1$, the load $W_t$ and the response time $Y_t$ are distributed as
\begin{align*}
W_t &\sim \mathrm{Beta}(2, 5), \quad\quad\quad\quad\quad\quad\quad\quad\quad\quad\text{for all }t,\\
Y_t &= 34.3 W_t + 250(C_t - 0.5)^2 + 250(M_t - 0.5)^2 + \\
              &\quad 200(C_t - 0.5)(M_t - 0.5), \quad\quad\quad\quad\text{ for all }t.
\end{align*}
In Scenario $2$, the system behaves as in Scenario $1$ up until time step $t=10$, after which the load is defined as
\begin{align*}
W_t &= \min\{1, 0.1 + 0.1(t-10)\}, &&\text{for }t > 10.
\end{align*}
This load function models a system under increasing stress (e.g., due to a gradual increase in UEs), which causes the safe operating region to slowly shrink over time.

Finally, we define the dynamics of the response time $Y_t$ as
\begin{align*}
  Y_t &= 34.3 W_t + 250(C_t - 0.5)^2 + 250(M_t - 0.5)^2 +\\
      &\quad 350\sin\left(\frac{t}{2}\right)\cdot (C_t - 0.5)(M_t - 0.5), \quad\quad\text{for }t > 10.
\end{align*}

\vspace{2mm}
\noindent\textit{\textbf{Experiment setup.}}
We implement COL in Python and run all simulations on a MacBook Pro with macOS Sequoia 15.6.1 and Python 3.11. For further implementation details, see Appendix \ref{app:experimental_setup}. We compare the performance of COL with two safe reinforcement learning algorithms: CPO \cite{pmlr-v70-achiam17a} and PPO-Lagrangian \cite{ray2019benchmarking}. We note that these algorithms have a different objective than COL. Rather than trying to learn as much as possible about the safe region, they aim to find a control policy that maximizes a reward function while satisfying the safety constraints. To enable a comparison with COL, we instantiate these algorithms with the constraint defined in \eqref{eq:example_specification} and define the reward at each time step to be $-C_t - M_t$, i.e., the reward function encodes the objective of minimizing resource usage.

\vspace{2mm}
\noindent\textit{\textbf{Simulation results.}}
The evolution of the safe region estimated through COL in simulation Scenario $1$ is illustrated in the upper plots of Fig.~\ref{fig:example_learning}. We see that the region inferred from observational data at time $t_0$ is a small subset of the safe region. Further, we observe that this subset gradually expands as interventions are carried out. Most interventions prescribed by COL are within the safe region, which is consistent with Prop.~\ref{prop:safe_confidence}. A few interventions are outside of the safe region, which can be explained by the confidence level $\alpha < 1$.

The evolution of the safe region estimated through COL in Scenario $2$ is illustrated in the lower plots of Fig.~\ref{fig:example_learning}. We observe that the safe region changes shape and orientation over time, which makes it difficult for COL to select safe interventions. As a consequence, the estimated safe region is smaller than in Scenario $1$. However, we also note that COL quickly adapts the estimated safe region as new observations are collected.

Lastly, Fig.~\ref{fig:safe_rl_2} and Table~\ref{tab:simulation_safe_rl} present a comparison of COL with the reinforcement learning baselines. We find that COL is several orders of magnitude more sample-efficient than the baselines. In Scenario~1, COL requires only $6$ unsafe interventions to converge, whereas CPO and PPO-Lagrangian require more than $60,000$ unsafe interventions to converge. 

\begin{figure}[H]
  \centering
  \scalebox{0.75}{
   \includegraphics{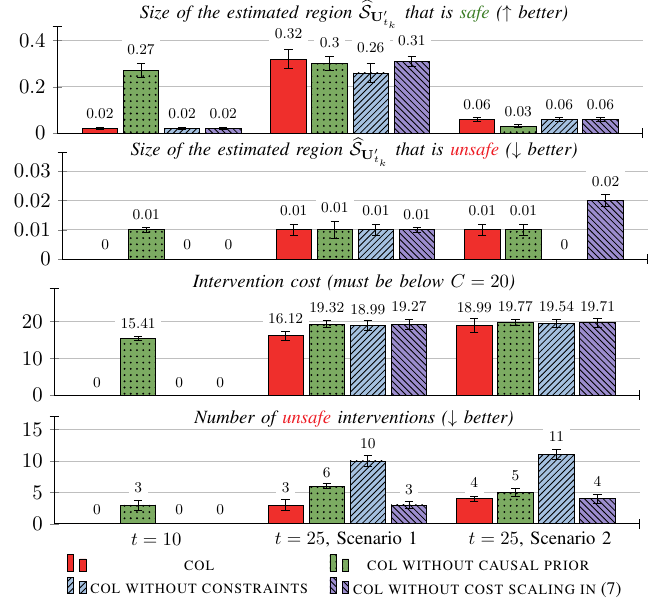}    
  }
  \caption{Ablation results for the example system in Fig.~\ref{fig:example_system}. Bars relate to different COL configurations; bar groups relate to time steps; numbers and error bars indicate the mean and the standard deviation from $10$ evaluations.}
  \label{fig:ablation_study}
\end{figure}

\vspace{2mm}
\noindent\textit{\textbf{Ablation study.}} To understand the relative importance of each component of COL, we evaluate COL with and without the causal prior \eqref{eq:initial_region}, the safety constraint \eqref{eq:safety_constraint}, and the denominator in \eqref{eq:policy}. The results are summarized in Fig.~\ref{fig:ablation_study}. We observe that the causal prior improves sample efficiency by reducing the number of unsafe interventions required to learn the safe region. Similarly, we see that removing safety constraint \eqref{eq:safety_constraint} leads to an increase in the number of unsafe interventions. Finally, we find that omitting the cost scaling in \eqref{eq:policy} has only a minor effect on the performance of COL for this system.

\section{Experimental Evaluation on a 5G Testbed}\label{experimental_evaluation}
In this section, we demonstrate how COL can be applied to learn the safe region of the cloud RAN in Fig.~\ref{fig:target_system}, which we deploy on a 5G testbed. We begin by describing our testbed implementation. Next, we describe how we applied COL to identify the safe region during the operation of the RAN. Lastly, we present and discuss the experimental findings.

\subsection{Testbed Implementation}\label{sec:testbed_impl}
\begin{figure}
  \centering
  \scalebox{0.48}{
   \includegraphics{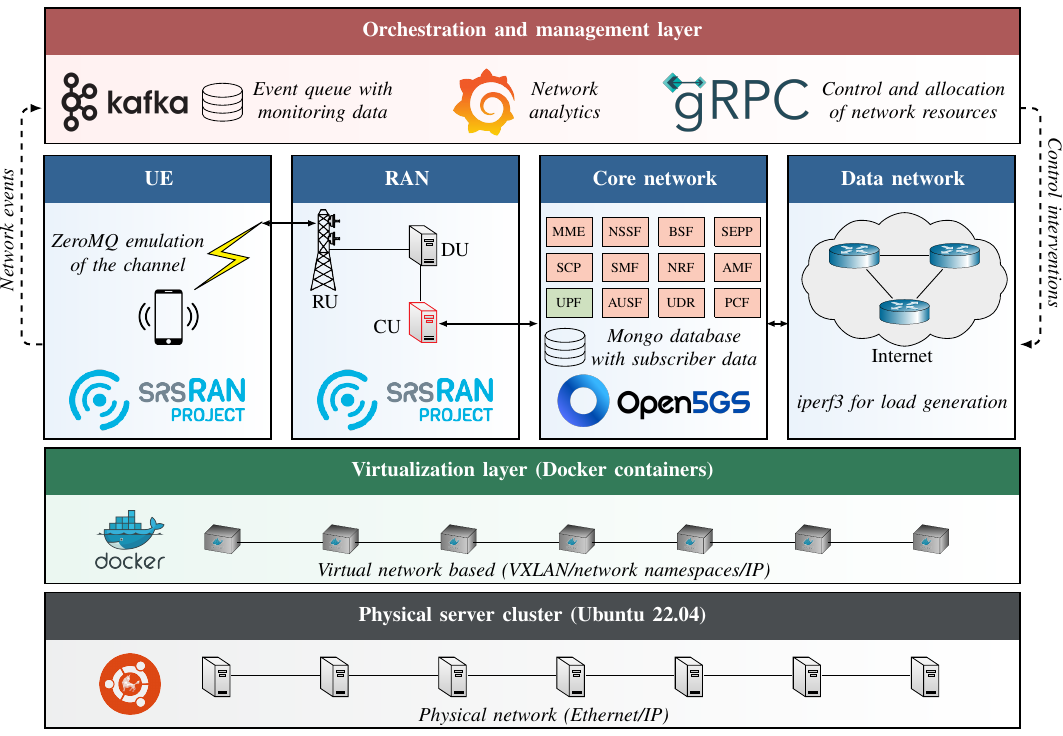}    
  }
\caption{Overview of our testbed that implements the 5G network shown in Fig.~\ref{fig:target_system}. It is deployed on a cluster of servers that run a virtualization layer based on Docker, which hosts the 5G network and the orchestration layer.}
  \label{fig:testbed}
\end{figure}

We implement the cloud RAN in Fig.~\ref{fig:target_system} on a testbed consisting of Supermicro 7049 servers that run Ubuntu 22.04; see Fig.~\ref{fig:testbed}. For the results reported in this paper, we use $2$ physical servers. Server 1 has a $24$-core Intel Xeon CPU and 768 GB RAM; server $2$ has a $16$-core Intel Xeon CPU and 128 GB RAM. They are interconnected via a 10 GbE link and run a virtualization layer provided by Docker Swarm and VXLAN. The RAN is deployed using the srsRAN 5G software-defined radio (SDR) stack \cite{srsran}, which provides software implementations of RUs, DUs, CUs, and UEs. Physical radio links are virtualized using ZeroMQ \cite{zeromq}. Finally, the core network is deployed as containerized services using Open5GS \cite{open5gs}, where network conditions on the fronthaul, midhaul, and backhaul links are created using NetEm in the Linux kernel \cite{netem}.

We configure the DUs to broadcast on a frequency channel in the 1800 MHz band with a bandwidth of 5 MHz. The RU connected to each DU uses a single-input single-output setup with one antenna, where the radio transceiver operates at a sampling rate of 5.76 MSps with transmission and reception gains set to 10 dB. All DUs, CUs, UEs, and services of the core network are allocated $20$ GB memory and $2.5$ CPU cores. We generate controlled uplink and downlink traffic using iperf3. In the downlink, traffic is sent from the core network toward the UE, while in the uplink, traffic is generated at the UE and transmitted through the RAN toward the core. 

\begin{figure}[H]
  \centering      
  \scalebox{0.7}{
   \includegraphics{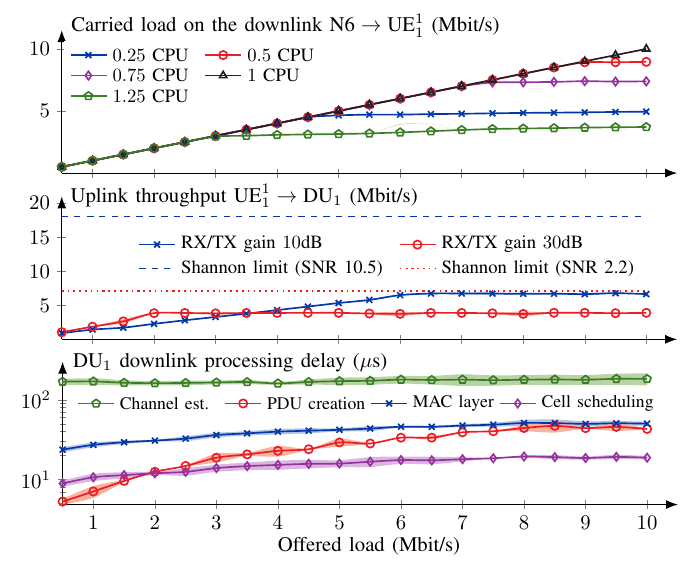}
  }
\caption{Testbed measurements of the 5G RAN in Fig.~\ref{fig:target_system} for varying uplink and downlink load, as indicated on the x-axes. Curves show the mean values based on five measurements; shaded regions indicate standard deviations.}\label{fig:monitoring_data}
\end{figure}

\begin{figure}
  \centering
  \scalebox{0.7}{
   \includegraphics{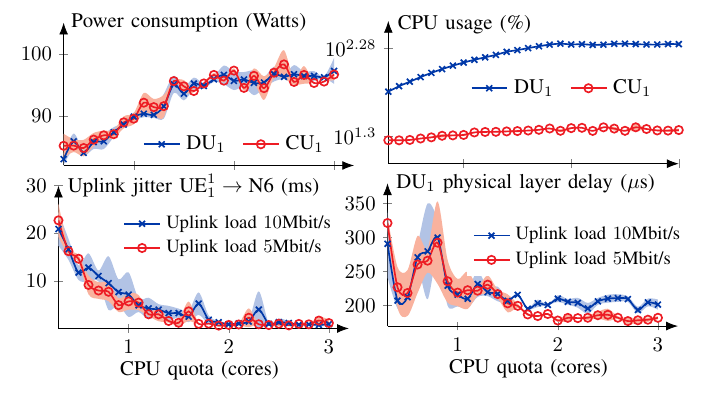}
  }
\caption{Testbed measurements of the 5G RAN shown in Fig.~\ref{fig:target_system} for varying CPU allocations, as indicated on the x-axes. Curves show the mean values based on five measurements; shaded regions indicate standard deviations.}\label{fig:monitoring_data_2}
\end{figure}

\vspace{2mm}
\noindent\textit{\textbf{Monitoring setup.}} We collect a broad set of performance metrics from the testbed, including all of the observable and target metrics in Fig.~\ref{fig:causal_structure_5g}. We aggregate these metrics over 15-second measurement windows and stream them to a Kafka backend for temporary storage and processing \cite{kafka}. Figures~\ref{fig:monitoring_data} and \ref{fig:monitoring_data_2} illustrate representative measurements collected from our testbed under varying traffic load and CPU allocations.

\subsection{Testbed Scenarios}\label{sec:testbed_scenarios}
To evaluate our method for learning the safe region of the RAN in Fig.~\ref{fig:target_system}, we consider the following evaluation scenarios.

\vspace{2mm}
\noindent\textit{\textbf{Scenario 1: CPU allocation under constant load}}. In this scenario, each DU is offered a constant load of 10 Mbit/s on both uplink and downlink. The goal is to learn the CPU allocations for the DUs and CUs that are safe in the sense that they ensure that the end-to-end throughput (downlink and uplink) is above $5$ Mbit/s and that the power consumption is below 95 Watts for a time horizon of $H=20$ time steps. We can express this safety specification formally as
\begin{align*}
  \varphi(\mathbf{Y}_{t_k}, \hdots,\mathbf{Y}_{t_k+20}) &= \square_{\tau \in [t_k,t_k+20]}\big(\Tilde{T}^i_{\tau} > 5 \land \Tilde{\widehat{T}}^i_{\tau} > 5 \land \\
  &\quad\quad\quad\quad\quad\quad\quad\Tilde{P}^i_{\tau} < 95 \land \Tilde{\mathscr{P}}^j_{\tau} < 95\big),
\end{align*}
for all DUs $i \in \{1,2,3,4\}$ and CUs $j \in \{1,2\}$, where the intervention set is $\mathbf{U}'_t=\{C^1_t, C^2_t, C^3_t, C^4_t, \mathscr{C}^1_t, \mathscr{C}^2_t\}$ and each CPU allocation is constrained to take values in the range $[0.3, 3]$ cores. The causal graph in Fig.~\ref{fig:causal_structure_5g} implies that the causal effects of such interventions are identifiable; cf.~Def.~\ref{def:ID}. (This can be verified numerically using the ID algorithm \cite{10.5555/2073876.2073938}.)

\vspace{2mm}
\noindent\textit{\textbf{Scenario 2: Traffic steering under dynamic load}}. In this scenario, the load is dynamic and determined by the number of UEs, which evolves according to the following dynamics. Initially, $5$ UEs are attached to each DU. At each time step, a new UE attaches to each DU with probability $p=0.15$ and departs with the same probability. Once a UE has attached, it offers a constant $0.75$ Mbit/s uplink and downlink load. 

We assume overlapping coverage between cells 1–2 and 3–4 in Fig.~\ref{fig:target_system}, which allows inter-frequency handovers. In particular, when the difference in uplink load between two neighboring cells exceeds $R_t \in [0,5]$, the cell with the highest load will hand over a UE to its neighbor. The goal is to learn the region of imbalance thresholds $R_t$ for which the resulting traffic steering decisions remain safe in the sense that load imbalances are mitigated without inducing excessive handovers. We formalize this notion of safety by requiring that the uplink jitter $J_t^i$ remains below $3.75$ ms over a horizon of length $H=5$, as expressed by the following specification.
\begin{align*}
\varphi(\mathbf{Y}_{t_k}, \hdots,\mathbf{Y}_{t_k+5}) &= \square_{\tau \in [t_k,t_{k+5}]}J^i_t < 3.75,
\end{align*}
for all DUs $i \in \{1,2,3,4\}$. It follows from the causal graph in Fig.~\ref{fig:causal_structure_5g} that the effect on this specification caused by changing the load-imbalance threshold $R_t$ is identifiable; cf.~Def.~\ref{def:ID}.

\begin{figure*}
  \centering      
  \scalebox{0.65}{
   \includegraphics{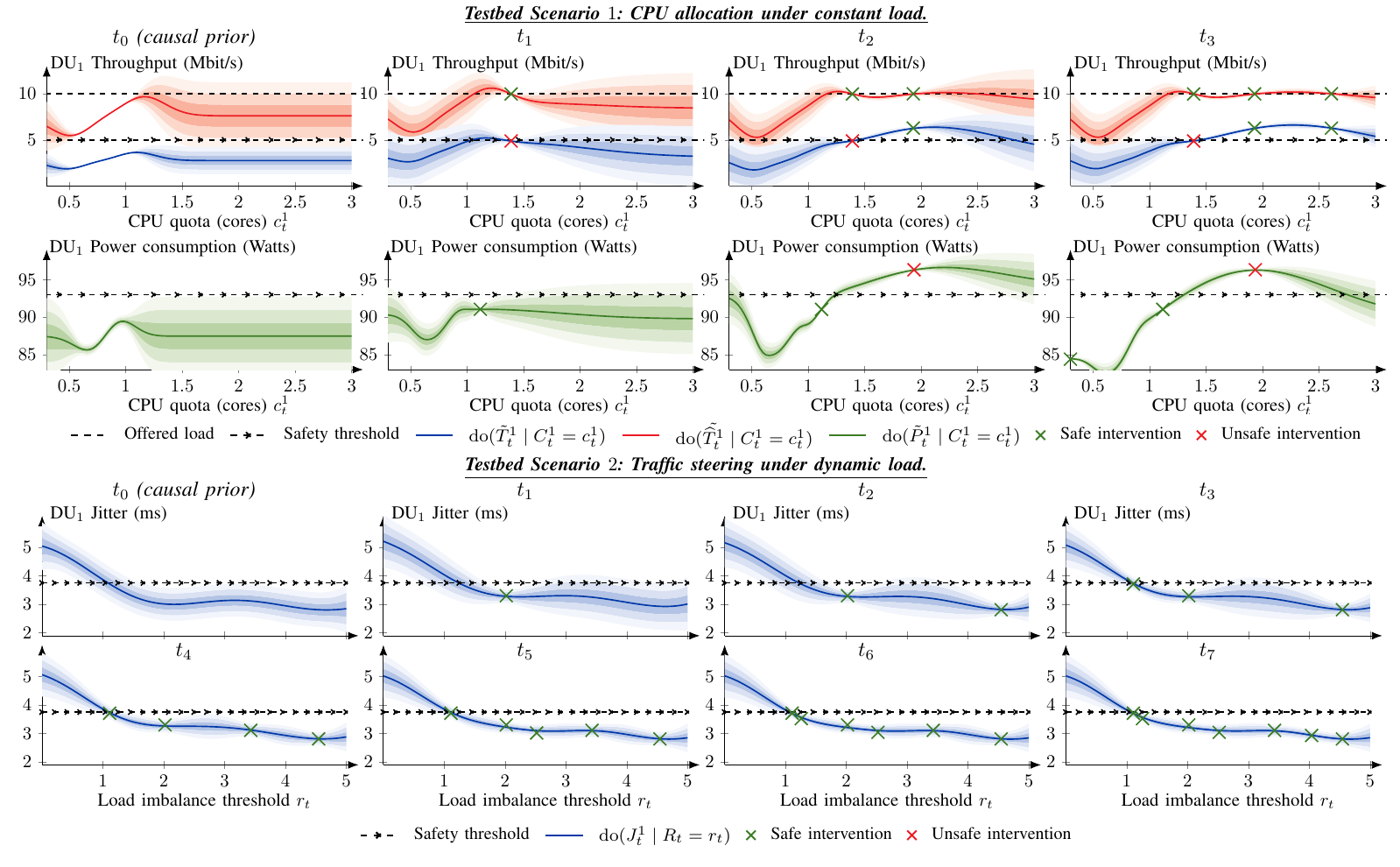}
  }
\caption{Causal effects estimated by COL [cf.~Alg.~\ref{alg:our_method}] on our testbed. The upper plots relate to Scenario $1$ and show how the estimated causal effects of the CPU allocation evolve from the causal prior at time $t_0$ to $t_3$. Similarly, the lower plots relate to Scenario $2$. They show how the estimated causal effects of the load imbalance threshold evolve from time $t_0$ to $t_7$. Curves show the mean values of the GPs; shaded regions indicate one, two, and three standard deviations from the mean (darker to lighter shades). The dotted and marked lines indicate the safety thresholds according to the specifications in \S\ref{sec:testbed_scenarios}.}\label{fig:causal_prior}
\end{figure*}

\subsection{Experiment Setup}
We instantiate COL with safety level $\delta=0.8$, confidence level $\alpha = 0.7$, and cost budget $C=100$. We define the cost of intervening on resource variables (e.g., the CPU allocation $C^i_{t}$ and memory allocation $M^i_{t}$) to be $1$ and the cost of intervening on the others (e.g., the load imbalance threshold $R_t$) to be $0.1$.

We follow the computational setup described in \S\ref{sec:ill_example}, with the key difference that the interventions selected by COL are executed on the testbed rather than in simulation. In particular, we apply the interventions to the testbed remotely via SSH and record the resulting observations through the monitoring infrastructure of the testbed. Due to the severe sample inefficiency of the safe reinforcement learning baselines [cf.~Fig.~\ref{fig:safe_rl_2}], it is practically infeasible to evaluate them on the testbed. For example, our simulation results indicate that CPO and PPO-Lagrangian require more than 60,000 interventions to converge. Considering that each intervention incurs a settling time of roughly 15 seconds, this would correspond to over 250 hours of continuous testbed operation (assuming $H=1$). For this reason, we cannot practically evaluate the safe reinforcement learning methods on the testbed. Instead, we use a version of COL without the causal prior as the main baseline.

\vspace{2mm}
\noindent\textit{\textbf{Initial system observations.}} Before starting the experimental evaluation of COL and the baselines, we collect $168$ hours of testbed measurements by passively observing the RAN while offering each DU a uniformly distributed load between $0$ and $10$ Mbit/s on both uplink and downlink. During this monitoring phase, CPU allocations for DUs and CUs are varied between $0.6$ and $1.0$ cores, signal gains are varied between $10$dB and 50dB, and the load imbalance threshold $R_t$ is varied between $0$ and $5$. All other variables are fixed to the values described in \S \ref{sec:testbed_impl}. We use the resulting system observations to define the causal prior for COL; see Lines 1 and 2 in Alg.~\ref{alg:our_method}.

\subsection{Evaluation Results}
\begin{figure}[H]
  \centering
  \scalebox{0.72}{
   \includegraphics{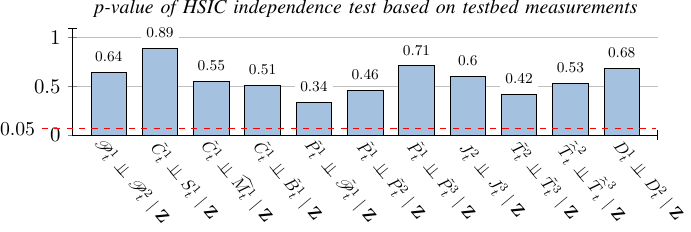}    
  }
  \caption{Validation of independence relations implied by the causal graph in Fig.~\ref{fig:causal_structure_5g} using testbed measurements. The y-axis shows p-values from HSIC independence tests \cite{hsic}; the dashed red line marks the $0.05$ significance level. Bars above the threshold are consistent with the graph. (Here $A \indep B \mid \mathbf{Z}$ denotes conditional independence given the separating set $\mathbf{Z}$ in the graph.)}
  \label{fig:causal_test}
\end{figure}

Figure~\ref{fig:causal_test} shows a validation of the causal graph in Fig.~\ref{fig:causal_structure_5g} using the testbed measurements. In particular, the figure shows the outcomes of statistical tests of the independence relations implied by the causal graph. We conduct these tests using the HSIC method \cite{hsic}. For all tested relations, we find that the p-values exceed the $0.05$ significance level, which indicates that the observed measurements are consistent with the independence assumptions encoded in the graph.

Figure~\ref{fig:causal_prior} shows how the causal effects estimated by COL evolve at the beginning of each evaluation scenario. At time $t_0$, COL is initialized with a causal prior learned from the initial monitoring data. We observe in the upper plots of Fig.~\ref{fig:causal_prior} that this prior assigns low uncertainty to the causal effects of CPU allocations in the range $[0.6,1.0]$ cores, which coincides with the range of allocations observed in the monitoring data. However, the prior exhibits high uncertainty for allocations outside this interval. As the evaluation proceeds, this uncertainty is reduced by forcing the CPU allocation to new values.

For Scenario~2, the lower plots in Fig.~\ref{fig:causal_prior} show the evolution of the estimated causal effects of the load-imbalance threshold $R_t$ on the jitter of the uplink of $\text{DU}_1$. At time $t_0$, the causal prior learned from the monitoring data already exhibits low uncertainty across the full range $R_t \in [0,5]$, which is covered in the monitoring data. As a result, COL requires fewer unsafe interventions in this scenario. In fact, all interventions shown in Fig.~\ref{fig:causal_prior} are confined to the safe region.

\begin{figure}
  \centering      
  \scalebox{0.69}{
   \includegraphics{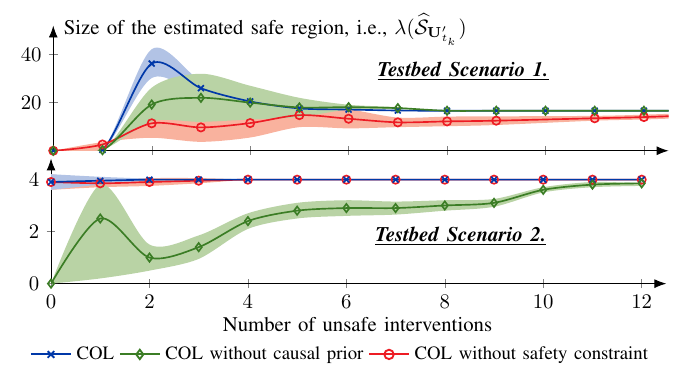}
  }
\caption{The size of the safe region estimated by COL [cf.~Alg.~\ref{alg:our_method}] and baseline methods on our testbed. Curves indicate the mean and the shaded areas indicate the standard deviation from $10$ evaluations.}\label{fig:learning_curves}
\end{figure}

Lastly, Fig.~\ref{fig:learning_curves} compares the performance of COL with and without the causal prior and safety constraint \eqref{eq:safety_constraint}. The figure reports the size of the estimated safe region as a function of the number of unsafe interventions. We observe in the upper plot that COL rapidly expands the safe region during the first interventions in Scenario~1 and then converges. Moreover, we see that removing the safety constraint results in more unsafe interventions. However, removing the causal prior has little impact in this scenario, which can be explained by the limited variability of the CPU allocations in the initial data. By contrast, the results in Scenario~2 (see the lower plot in Fig.~\ref{fig:learning_curves}) show that the causal prior provides an accurate estimate of the safe region, which reduces the need for unsafe interventions and leads to around $10\times$ higher sample efficiency.

\vspace{2mm}
\noindent\textit{\textbf{Takeaways.}} In summary, our main experimental findings are:
\begin{itemize}
  \item \textit{Causal structure enhances safety and sample efficiency.}
\begin{itemize}
\item By utilizing the causal graph (validated in Fig.~\ref{fig:causal_test}), COL can safely infer the effects of identifiable interventions without having to intervene on the RAN. These inferences allow us to establish a \textit{causal prior} of the safe region [cf.~\eqref{eq:initial_region}], which significantly reduces the number of unsafe interventions required to learn the safe region and thus improves sample efficiency; cf. the green and blue curves in Fig.~\ref{fig:learning_curves}.
\end{itemize}
\item \textit{Monitoring under varying conditions is important.}
\begin{itemize}
\item The effectiveness of the causal inference depends on the variability in the data. In the traffic steering scenario, the monitoring data covers the full range of the load-imbalance threshold. As a result, we obtain an accurate causal prior, requiring almost no unsafe interventions to identify the safe region; see the lower plots in Fig.~\ref{fig:causal_prior}. Conversely, in the CPU allocation scenario, the monitoring data is limited to a narrow range of CPU allocations ($0.6$–$1.0$ cores), which means that several unsafe interventions are required to accurately learn the safe region; see Fig.~\ref{fig:causal_prior}.
\end{itemize}
\item \textit{Intervention policy \eqref{eq:policy} enables safe active learning.}
\begin{itemize}
\item By selecting interventions based on the ratio of expected information gain to cost, policy \eqref{eq:policy} reduces uncertainty regarding the boundaries of the safe region while maintaining safety with high probability, as demonstrated in Fig.~\ref{fig:example_learning}. Moreover, we observe in Fig.~\ref{fig:ablation_study} and Fig.~\ref{fig:learning_curves} that removing constraint \eqref{eq:safety_constraint} significantly reduces the safety of the policy.
\end{itemize}
\end{itemize}
\section{Discussion On The Practicality of COL}
The main assumption underpinning COL is that the causal graph of the RAN is known.\footnote{COL can be applied in settings where the causal graph is unknown. In such cases, COL has lower sample efficiency and weaker theoretical guarantees.} We acknowledge that this can be a strong assumption, particularly for large-scale networks where causal dependencies may be opaque. However, for most RANs in production, network operators have detailed domain knowledge about the hardware and software architecture of the network, from which the causal graph can be constructed given certain engineering effort and time. Moreover, our experimental results demonstrate that this upfront engineering effort is justified as it significantly reduces the number of interventions required to accurately learn the safe region.

While this paper focuses on scenarios where the causal structure of the RAN is fixed, we note that the causal structure may also evolve over time, e.g., due to the deployment of new DUs and CUs. However, such structural changes in the network typically occur on a longer time scale than the RAN's operational dynamics. As a consequence, when such structural changes occur, the causal graph can be manually updated by operators during maintenance windows or automatically refined through causal discovery techniques; see e.g., \cite{NEURIPS2022_c9fcd02e}.
\subsection{Operational Considerations}
COL is designed for \textit{online} learning of the safe region and can be deployed either in a digital twin (i.e., a testbed \cite{10154288}) or directly in a production RAN. When used in a digital twin, interventions prescribed by COL can be safely applied without incurring operational costs or impacting production traffic and services. By contrast, when COL is deployed in a production environment, the operational cost is determined by the number of interventions and their impact on the RAN's performance and compliance with the specification. This cost can be controlled by tuning the confidence level $\alpha$ and the cost function $c$ when instantiating COL, as defined in Alg.~\ref{alg:our_method}. In particular, specifying the confidence level to be $\alpha = 1$ prevents any unsafe interventions, but may also slow down the learning of the safe region. Moreover, interventions with high operational impact can be assigned higher costs by the cost function $c$, which discourages COL from selecting them. 

\subsection{Use Cases of the Learned Safe Region}
Once the safe operating region has been identified with sufficiently high confidence through COL, it can be used to safely deploy learning-based control methods (e.g., reinforcement learning methods) for optimizing various control functions in the RAN.  Examples of such control functions include traffic steering \cite{10102369}, resource allocation \cite{9806318, 9705104}, task offloading \cite{9253665,8771176}, load balancing \cite{8334688}, service orchestration \cite{8924682}, incident response \cite{hammar2025incidentresponseplanningusing}, and failure recovery \cite{9303420,tifs_25_HLALB,dsn24_hammar_stadler}. In each case, the safe region restricts the admissible control inputs, allowing automated controllers to optimize the RAN's operation while avoiding configurations that could violate constraints. 

\subsection{Applications of COL Beyond Cloud RANs}
While we focus on cloud RANs in this paper, COL is a general method for safe online learning in networked systems. Its applicability extends to any system equipped with monitoring and control mechanisms, such as cyber-physical systems, enterprise networks, and general cloud infrastructures. Importantly, COL is not limited to a specific causal graph and can operate with both discrete and continuous system variables.

\section{Conclusion}
The flexibility of cloud radio access networks (RANs) creates new opportunities for autonomous network management through learning-based control. However, deploying such approaches in operational networks is challenging because they generally cannot ensure that service requirements are met during the learning process. To mitigate this challenge and enable reliable automation and control, we must identify the RAN's \textit{safe operating region}, i.e., the set of control inputs for which the RAN's specification is fulfilled. This region can then be used as a guardrail for automatic controllers, allowing them to safely optimize the RAN to meet management objectives.

In this paper, we present \textbf{C}ausal \textbf{O}nline \textbf{L}earning (COL), a novel method for identifying the safe operating region of a cloud RAN. By exploiting the causal structure of the RAN, COL combines causal inference with active Bayesian learning to safely and efficiently learn the safe region during operation. We prove that COL provides probabilistic safety guarantees throughout learning and converges to the full safe region under standard conditions. Experiments on a 5G testbed demonstrate that COL reliably learns the safe region and is up to $10\times$ more sample-efficient than state-of-the-art methods for safe learning. A key lesson learned from our experiments is that the effort required to model the RAN’s causal structure is justified, as it enables reliable and efficient identification of the safe region.

\vspace{2mm}
\noindent\textit{\textbf{Future work.}} We plan to leverage the safe region learned by COL to implement safe automation and optimization of control functions in the RAN, such as load balancing \cite{8334688}. In addition, we plan to extend COL to settings where the causal graph is unknown by integrating causal discovery techniques that allow us to infer the graph from system observations \cite{10.5555/1597348.1597382}.

\section*{Acknowledgments}
This research is supported by the Swedish Research Council under contract 2024-06436. The authors want to thank the RAN autonomy group at Telstra for constructive discussions related to the use case investigated in this work. (Telstra operates the largest RAN in Australia.)
\appendices

\section{Proof of Proposition~\ref{prop:safe_confidence}}\label{app:gp_bound}
We define the constant $\kappa_t(\mathbf{u}')$ in \eqref{eq:region_estimate} as
\begin{align*}
\kappa_t(\mathbf{u}') &= \beta^{1/2}_t\sqrt[]{k_{\mathcal{D}_t}(\mathbf{u}',\mathbf{u}')}.
\end{align*}
We assume that the sampling noise is uniformly bounded by $\sigma$ and define the constant $\beta_t$ as 
\begin{align*}
\beta_{t} &= 2B^2 + 300\gamma_t \ln^3(t/(1-\alpha))
\end{align*}
where $\alpha \in (0,1)$ is a configurable constant,
\begin{align*}
\gamma_{t} &= \max_{\mathcal{D}: |\mathcal{D}|=t} I(\mathcal{D}; p_{\mathbf{U}'_t}(\mathbf{u}'_t)),
\end{align*}
and $B$ is an upper bound on the RKHS norm, i.e.,
\begin{align*}
\norm{p_{\mathbf{U}'_t}(\mathbf{u}'_t)}_{\mathrm{RKHS}} \leq B,
\end{align*}
Here $\mathcal{D}_{t}$ is the set of system trajectories and $I(\mathcal{D}; p_{\mathbf{U}'_t}(\mathbf{u}'_t)\})$ denotes the reduction in uncertainty about the causal effect $p_{\mathbf{U}'_t}(\mathbf{u}')$ when revealing the observations in the dataset $\mathcal{D}$, i.e., the information gain. This gain is given by
\begin{align*}
\frac{1}{2} \ln|\mathbf{I} + \sigma^{-2}[k(\mathbf{u}',\mathbf{u}'')]_{\mathbf{u}',\mathbf{u}'' \in \mathcal{D}}|,
\end{align*}
where $\mathbf{I}$ is the identity matrix. Condition a) of Prop.~\ref{prop:consistency} implies that the GP is well-defined. Moreover, as is well-known, the above conditions imply that the following inequality holds
\begin{align}
P(|p_{\mathbf{U}'_t}(\mathbf{u}')\!\!-\!\! m_{\mathbf{U}'_t|\mathcal{D}_t}(\mathbf{u}')| \leq \kappa_t(\mathbf{u}')) \geq \alpha,\label{eq:gp_bound}
\end{align}
as proven by Srinivas et al. \cite[Theorem 6]{gp_ucb}. This inequality implies that the estimated region $\hat{\mathcal{S}}_{\mathbf{U}'_t}$ [cf.~\eqref{eq:region_estimate}] will only include interventions that are safe with probability at least $\alpha$. \qed

\section{Hyperparameters and Implementation Details}\label{app:experimental_setup}
We implement the GPs using the BoTorch library~\cite{10.5555/3495724.3497531} with default hyperparameters. We implement the causal inference logic using the DoWhy library~\cite{sharma2020dowhyendtoendlibrarycausal}. Finally, we implement the safe reinforcement learning baselines using the OmniSafe framework~\cite{ji2023omnisafeinfrastructureacceleratingsafe} with default hyperparameters, except for the initial Lagrange multiplier, which we set to $0.55$. Our software implementations are open source and available at \cite{csle_docs}.

\bibliographystyle{IEEEtran}
\bibliography{references}

\end{document}